\documentclass[aps,pre,onecolumn,showpacs,superscriptaddress,noeprint]{revtex4-1}
\usepackage[utf8]{inputenc}
\usepackage{amsmath}
\usepackage{multirow,array}
\usepackage[margin=0.9in]{geometry}
\usepackage{graphics}
\usepackage{graphicx}
\usepackage{subfigure}
\usepackage{color}

\begin{document}

\title{Asymmetric games on networks: mapping to Ising models and bounded rationality} 
\author{Filippo Zimmaro}
\email{zimmarofilippo@gmail.com}
\affiliation{Department of Mathematics, University of Bologna, Bologna, Italy}
\affiliation{Department of Computer Science, University of Pisa, Pisa, Italy}

\author{Serge Galam}
\affiliation{CEVIPOF - Centre for Political Research, Sciences Po and CNRS, Paris, France}

\author{Marco Alberto Javarone}
\email{marcojavarone@gmail.com}
\affiliation{Dipartimento di Fisica, Università di Bari, Bari, Italy}
\affiliation{Dutch Institute for Emergence Phenomena, Amsterdam, Netherlands}  

\date{\today}

\begin{abstract} 
We investigate the dynamics of coordination and consensus in an agent population. Considering agents endowed with bounded rationality, we study asymmetric coordination games using a mapping to random field Ising models. In doing so, we investigate the relationship between group coordination and agent rationality. Analytical calculations and numerical simulations of the proposed model lead to novel insight into opinion dynamics. For instance, we find that bounded rationality and preference intensity can determine a series of possible scenarios with different levels of opinion polarization. To conclude, we deem our investigation opens a new avenue for studying game dynamics through methods of statistical physics.
\end{abstract}

\maketitle
\section{Introduction}\label{sec:introduction}
\noindent
Group coordination is a collective phenomenon of particular interest in various contexts~\cite{read01} and can be observed in human communities and other animal groups.
For instance, flocks, schools of fish, and ant colonies often show hallmarks of group coordination. The latter emerges, typically, to address specific functions, such as improving the quality of flights and defending from attacks~\cite{corcoran01}. In human societies, to cite a few, coordination underlies several activities, such as sports, business developments, start-up growth, and company organisation.
Therefore, understanding group coordination and its mechanisms can clarify relevant social aspects of our society and nature.
To this end, game theory allows mapping group coordination to a specific equilibrium in a competition among strategies.
For instance, let us consider a group of friends conversing about music genres, sports teams, or political candidates. Here, mapping opinions to strategies and exchanges of ideas to game interactions, the convergence of the group to a common opinion corresponds to the success of a strategy. 
These friends, mapped to agents playing a game, may have personal preferences. So, in the presence of an opinion of the majority, we wonder whether agents biased towards a different opinion can suffer from social pressure.
On the one hand, falsifying their preference~\cite{siegel01}, i.e. avoiding expressing an opposite opinion, may reduce conflicts and lead to group coordination. On the other hand, in some cases, declaring an honest view can be more profitable.
In summary, the described real-world scenario, i.e. the conversating friends, mapped to an agent population allows us to exploit game theory. For instance, in game theory, agents are defined as rational when they act to maximise their payoff and can undergo a strategy revision phase~\cite{javarone01,perc01}, so the resulting evolutionary dynamics can lead the population towards some strategy equilibria corresponding to opinion consensus/dissent. Notice that combining game theory with evolutionary mechanisms is at the core of the evolutionary game theory framework~\cite{nowak01,szolnoki00,duh01,santos01,szolnoki02,szolnoki03,javarone02}, fundamental for studying the strategy equilibria in various scenarios.\\
\\
We consider a population whose agents interact by playing asymmetric games. Thus, conflicting preferences motivate the emergence of equilibria where group coordination is not reached. Yet, we aim to quantify these mechanisms and measure the polarisation under different conditions, such as various levels of rationality. 
Concerning that, previous studies focused on similar aspects. To cite a few, \cite{hernandez2013heterogeneous,hernandez2017equilibrium,broere2017network,peidynamic} study the asymmetric games theoretically and through simulations assuming best-response dynamics,~\cite{broere2019experimental,goyal2021integration} perform experiments with human subjects on different social networks. \cite{galam1997rational,galam2008sociophysics} study a version of the random field Ising model at temperature $T=0$, while~\cite{galam2010ising} focuses on the mapping between a general 2-players game and the Ising model. In~\cite{correia2022asymmetric}, the authors attempt to connect the asymmetric games on the network to an Ising model, which is encouraged in the review \cite{broere2020essays}.
Our main contributions include identifying the conditions for which the studied asymmetric games are potential games \cite{szabo2016evolutionary} by mapping them to a random field Ising model and, through the latter, studying analytically the effects of bounded rationality in asymmetric games on networks.
We remark that random fields have already been investigated within the context of social dynamics~\cite{galam1991towards} to describe the dynamics of opinion consensus in open and closed (i.e. finite size) populations.
Then, our results, supported by numerical simulations, show a rich spectrum of outcomes leading to various interpretations, such as opinion polarization. The latter becomes particularly interesting as it relates to the bounded rationality assumption considered in the proposed model.\\
\\
The remainder of the manuscript is organised as follows. In Section~\ref{sec:model}, we introduce asymmetric games on networks with more detail, give a condition for them to be potential games, and map their dynamics to an Ising model. Then, in Section~\ref{sec:results}, we study the model on a complete network and a $k$-regular network, while Section~\ref{sec:results} focuses more on numerical simulations. Finally, we discuss the main finding in Section~\ref{sec:conclusion}. Two additional appendices report, respectively, the calculations to map the game dynamics to the Ising model (Appendix ~\ref{sec:general_mapping}) and the study of asymmetric games with infinite rationality (Appendix ~\ref{sec:best_response}).

\section{Model}\label{sec:model}
\noindent
Let us now introduce the asymmetric game we use on networks to study the dynamics of group coordination, namely the $2$-player Battle of Sexes (BoS) \cite{hofbauer1998evolutionary}. This game has two strategies, $x_i=\{0,1\}$, and one fixed identity $\theta_i = \{0,1\}$. The latter identifies the strategy preference of each player, e.g. $\theta_i = 1$ indicates that the $i$-player prefers the strategy $1$ (and vice versa). 
So, the possible combinations of players' identities lead to the following payoff matrices
\begin{equation}
\begin{array}{cc|c|c|}
      & \multicolumn{1}{c}{} & \multicolumn{2}{c}{\mathbf{\theta_j = 1}}\\
      & \multicolumn{1}{c}{} & \multicolumn{1}{c}{1}  & \multicolumn{1}{c}{0} \\\cline{3-4}
     \multirow{2}{*}{$\mathbf{\theta_i=1}$}  & 1 & (1,1) & (0,0) \\ \cline{3-4}
      & 0 & (0,0) & (S,S) \\ \cline{3-4}
\end{array}
\quad
\hspace{0.4cm}
\begin{array}{cc|c|c|}
      & \multicolumn{1}{c}{} & \multicolumn{2}{c}{\mathbf{\theta_j = 0}}\\
      & \multicolumn{1}{c}{} & \multicolumn{1}{c}{1}  & \multicolumn{1}{c}{0} \\\cline{3-4}
      \multirow{2}{*}{$\mathbf{\theta_i=1}$}  & 1 & (1,S) & (0,0) \\ \cline{3-4}
      & 0 & (0,0) & (S,1) \\\cline{3-4}
\end{array}
\quad
\hspace{0.4cm}
\begin{array}{cc|c|c|}
      & \multicolumn{1}{c}{} & \multicolumn{2}{c}{\mathbf{\theta_j = 0}}\\
      & \multicolumn{1}{c}{} & \multicolumn{1}{c}{1}  & \multicolumn{1}{c}{0} \\\cline{3-4}
      \multirow{2}{*}{$\mathbf{\theta_i=0}$} & 1 & (S,S) & (0,0) \\ \cline{3-4}
      & 0 & (0,0) & (1,1) \\\cline{3-4}
\end{array}
\label{pot_mat_B}
\end{equation}
\noindent where the elements of the matrices $(\cdot,\cdot)$ indicate respectively the reward of player $i$ and $j$ with the corresponding combination of strategies. $S\in[0,1]$ denotes a parameter related to the preference strength: the higher $S$, the lower the difference in terms of reward between coordination at the preferred and unpreferred strategy, so the lower the preference intensity. 
In a network, each agent plays the 2-players BoS with each one of its neighbours simultaneously, i.e. the agent's chosen strategy is the same for all its interactions \cite{broere2017network, hernandez2013heterogeneous}.
We can write each player's utility function (total payoff) as
\begin{equation}
    \pi_i(x_i;\theta_i) = \chi(x_i;\theta_i) \sum_{j\in\partial i} I_{\{x_i=x_j\}}
\end{equation}
where 
\begin{equation}
    \chi(x_i;\theta_i) = 
    \begin{cases}
        1 \;\;\;\; if \;\;\;\; x_i=\theta_i \\
        S \;\;\;\; if \;\;\;\; x_i\neq\theta_i \\ 
    \end{cases}
\end{equation}
\noindent and $I_{\{\cdot\}}$ is the indicator function, equal to $1$ if the condition $\cdot$ is verified and zero otherwise.
We refer to this class of games as the Broere's model \cite{broere2017network} and study the evolutionary dynamics of the agent population, considering various initial conditions. At each time step, an agent $i$ is randomly selected and chooses its strategy in function of the current configuration of its ego network (i.e. its neighbours) through the so-called Logit rule \cite{szabo2007evolutionary,perc2017statistical}. 
Specifically, the selected agent plays $x_i=1$ with probability 
\begin{equation}
    P_i(1) = \frac{e^{R_i[\pi_i(1;\theta_i)]}}{e^{R_i[\pi_i(1;\theta_i)]} + e^{R_i[\pi_i(0;\theta_i)}]} = \frac{e^{R_i[\pi_i(1;\theta_i) - \pi_i(0;\theta_i)]} }{1+e^{R_i[\pi_i(1;\theta_i) - \pi_i(0;\theta_i)]}}
\end{equation}
\noindent and $x_i=0$ with probability $P_i(0) = 1 - P_i(1)$. 
Notice that the probabilities depend only on the payoff difference between the two strategies. Also, we assume the agents have complete information about their ego networks. The term $R_i\in [0,+\infty)$ represents the rationality of the $i$-th agent and reflects the inclination to pursue its personal interest (i.e. maximizing its utility), but can have alternative interpretations later discussed. Accordingly, $R_i\rightarrow \infty$ entails the $i$-th agent playing the best response, whereas $R_i=0$ entails an irrational attitude as the strategy is randomly selected.
In general, we consider $R_i=R \;\forall i=1,...,N$, and we rescale the rationality parameter $R$ by a factor equal to the average degree of the network so that $R=\frac{r}{<k>}$.  \\
\\
Following the above prescriptions, let us consider, for example, the $i$-th agent with $k_i$ degree (i.e. number of neighbours) and identity $\theta_i=1$. By indicating with $w_i$ the number of $i$-th agent's neighbours currently playing $1$, $\pi_i(1;1) = w_i$ and $\pi_i(0;1) = S(k_i-w_i)$, so our agent chooses the strategy $1$ with probability $P_i(1) =  \frac{e^{R_i[w_i-S(k_i-w_i)]} }{1+e^{R_i[w_i-S(k_i-w_i)]}}$ and, indeed, the strategy $0$ with probability $1-P_i(1)$.

\subsection{Mapping to Ising models} \label{subsec:mapping}
In the appendix \ref{sec:general_mapping}, we show that a game on a network $G=(V,E)$ with payoff matrices of the type 
\begin{equation*}
\begin{array}{cc|c|c|}
      & \multicolumn{1}{c}{} & \multicolumn{2}{c}{\mathbf{\cdot}}\\
      & \multicolumn{1}{c}{} & \multicolumn{1}{c}{1}  & \multicolumn{1}{c}{0} \\\cline{3-4}
      \multirow{2}{*}{$\mathbf{i}$}  & 1 & (a^{(11)}_i,\cdot) & (a^{(10)}_i,\cdot) \\ \cline{3-4}
      & 0 & (a^{(01)}_i,\cdot) & (a^{(00)}_i,\cdot) \\\cline{3-4}
\end{array}
\end{equation*}
\noindent has a potential if and only if 
\begin{equation}
    (a_i^{(11)}+a_i^{(00)}) - (a_i^{(01)}+a_i^{(10)}) = C\;\;\;\;\;\forall i=1,...,N
\label{condition_pot}
\end{equation}
\noindent where $C$ is a constant.\\
Thus, assuming a homogeneous rationality $R$ and Logit rule underlying the system evolution, the game is equivalent to an Ising model evolving by the Glauber dynamics with Hamiltonian
$H = -\sum_{i\in V} h_i\sigma_i - J\sum_{ij\in E}\sigma_i\sigma_j$, where the strategies $\boldsymbol{x} =\{0,1\}^N$ are mapped to the spin random variables $\boldsymbol{\sigma} = \{-1,+1\}^N$. The mapping between the two models is realised via the following  correspondences:
\begin{equation}
\begin{split}
    &h_i = \frac{(a_i^{(11)}+a_i^{(10)}) - (a_i^{(00)}+a_i^{(01)})}{4} k_i \;\;\;\;\;\;\;\;\;
    \;\;\;\;\;\;\;\;\;i=1,...,N\\
    &J  = \frac{(a_i^{(11)}+a_i^{(00)}) - (a_i^{(01)}+a_i^{(10)})}{4} \\
    &\beta = R
\end{split}
\end{equation}
\noindent where $k_i$ is the degree of the $i$-th node in the network.\\
Now, applying these results to the payoff matrices (\ref{pot_mat_B}) associated with the Broere's model, we see that the payoff matrices can be of two types, depending on the individuals' preference. Identifying the agents with personal preference for the strategy $1$ (resp. $0$) as belonging to class $A$ (resp. $B$), we have that 
\begin{equation}
    \begin{split}
        a_A^{(11)}=1 \quad \;\; a_A^{(00)}=S \quad \;\;  a_A^{(01)}=a_A^{(10)}=0 \\
        a_B^{(11)}=S\quad \;\;  a_B^{(00)}=1 \quad \;\;  a_B^{(01)}=a_B^{(10)}=0
    \end{split}
\end{equation}
Being 
\begin{equation}
    (a_A^{(11)}+a_A^{(00)}) - (a_A^{(01)}+a_A^{(10)}) = (a_B^{(11)}+a_B^{(00)}) - (a_B^{(01)}+a_B^{(10)}) = 1+S
\end{equation}
the condition (\ref{condition_pot}) is satisfied and the game has a defined potential. Moreover, the parameters of the corresponding Ising model read
\begin{equation}
\begin{split}
    &h_i  =  \frac{1-S}{4} k_i \quad \quad \forall \;i\in A\\
    &h_j  =  \frac{S-1}{4} k_j \quad \quad \forall \;j\in B\\
    &J = \frac{1+S}{4} \\
    &\beta = R
\label{mapping general}
\end{split}
\end{equation}

\section{Results}\label{sec:results}
\subsection{Complete networks}
\begin{figure}[htbp]
  \centering
  \includegraphics[width=0.45\textwidth]{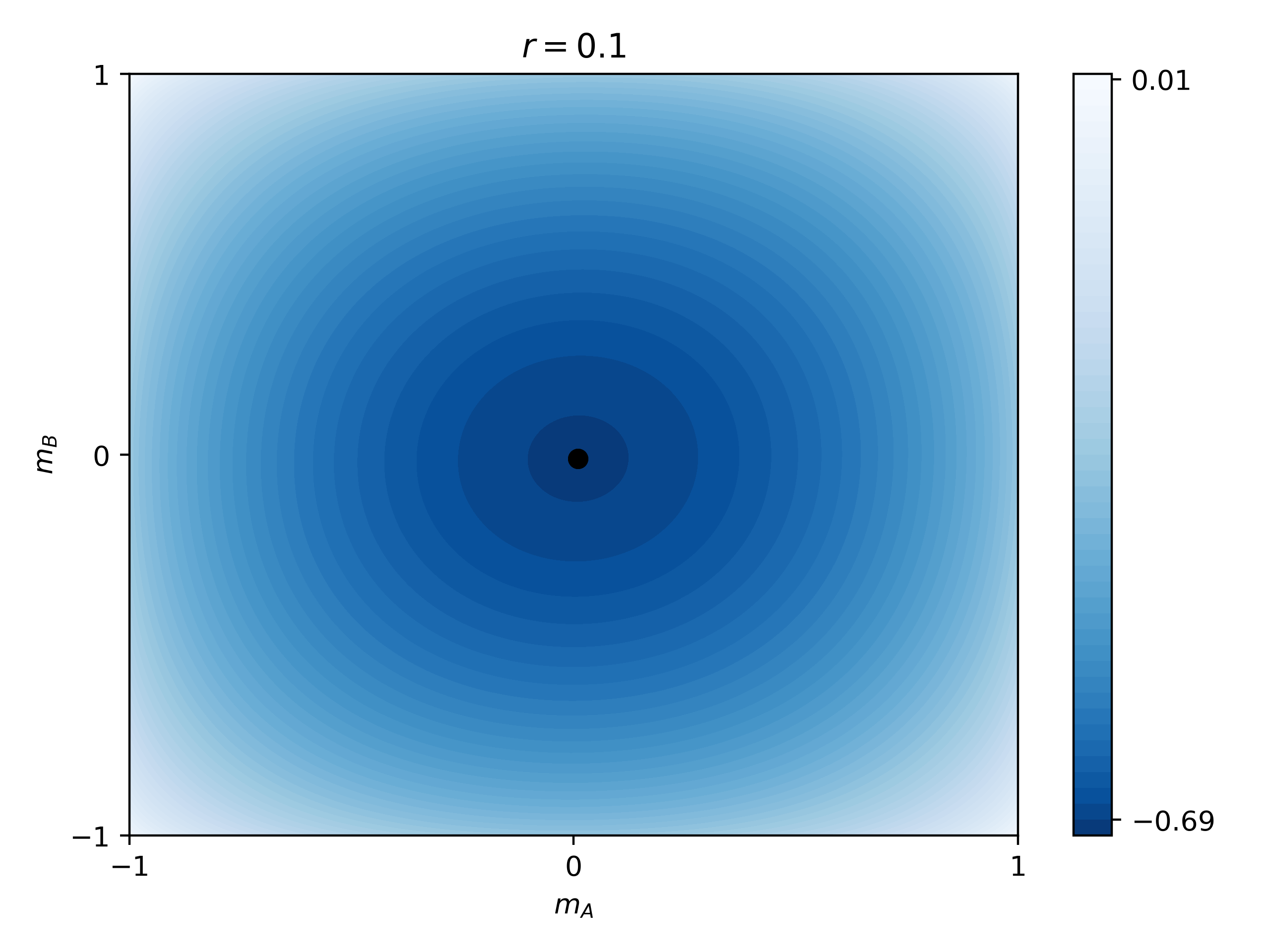}
  \includegraphics[width=0.45\textwidth]{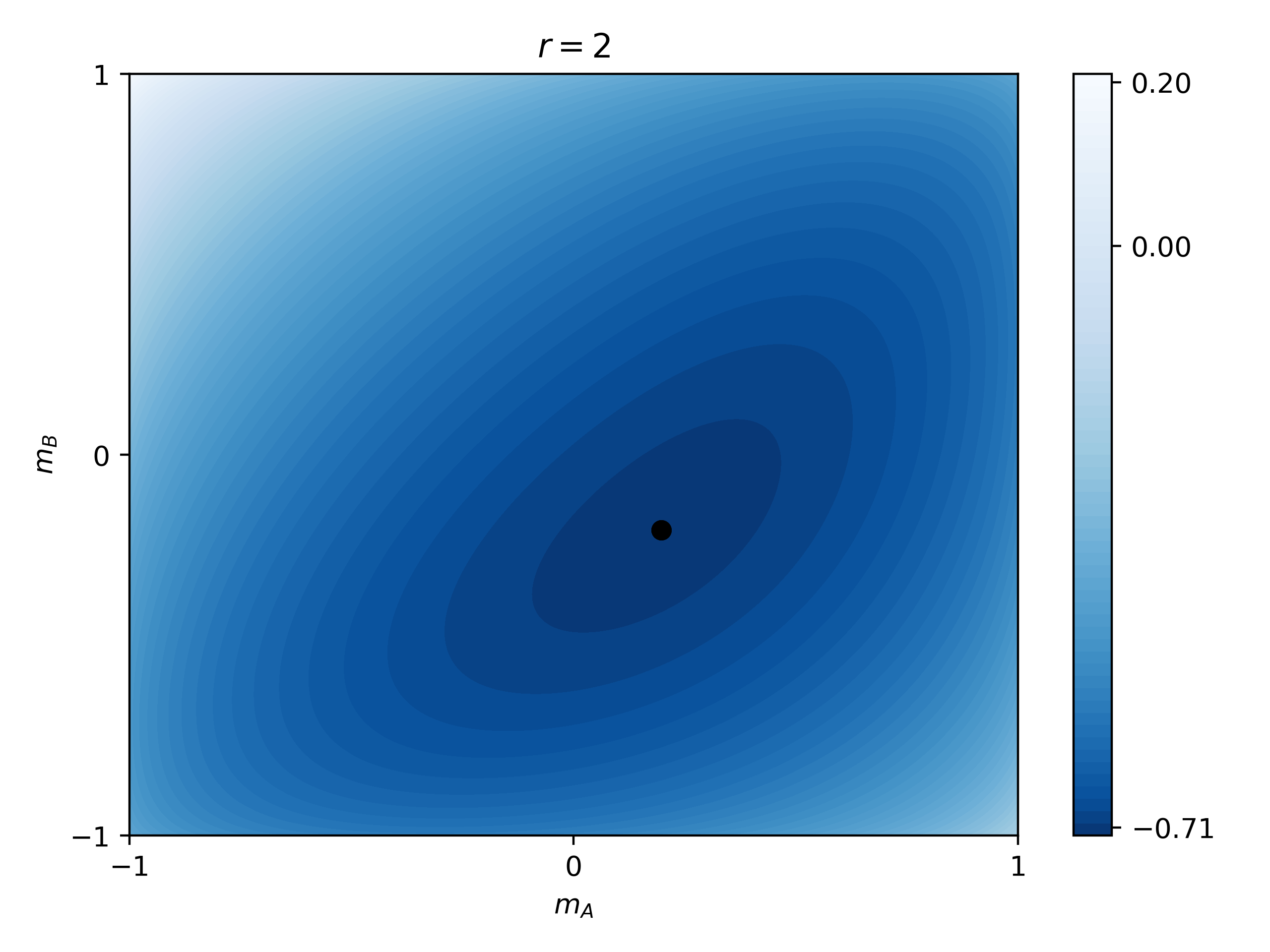}
  \includegraphics[width=0.45\textwidth]{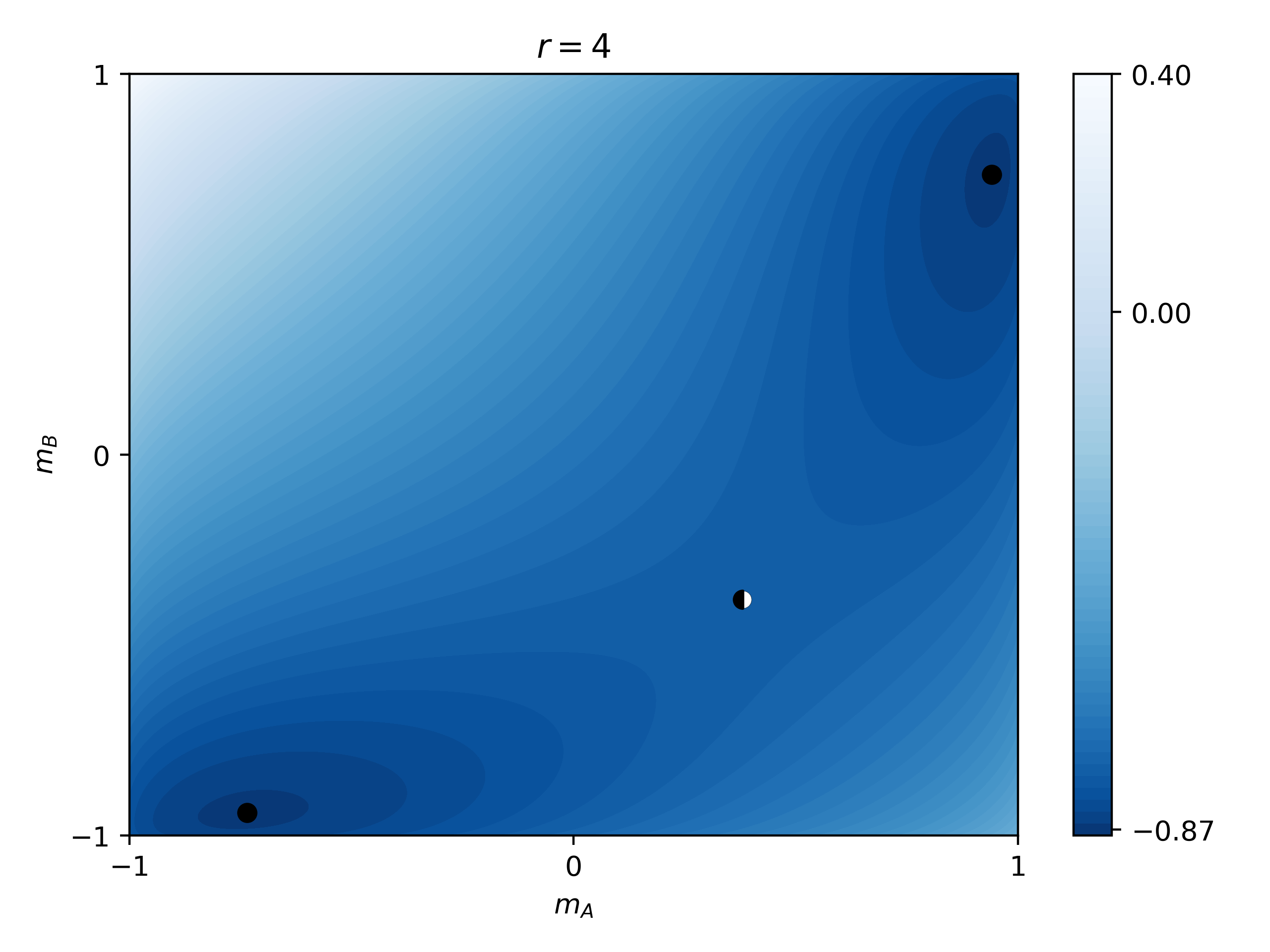}
  \includegraphics[width=0.45\textwidth]{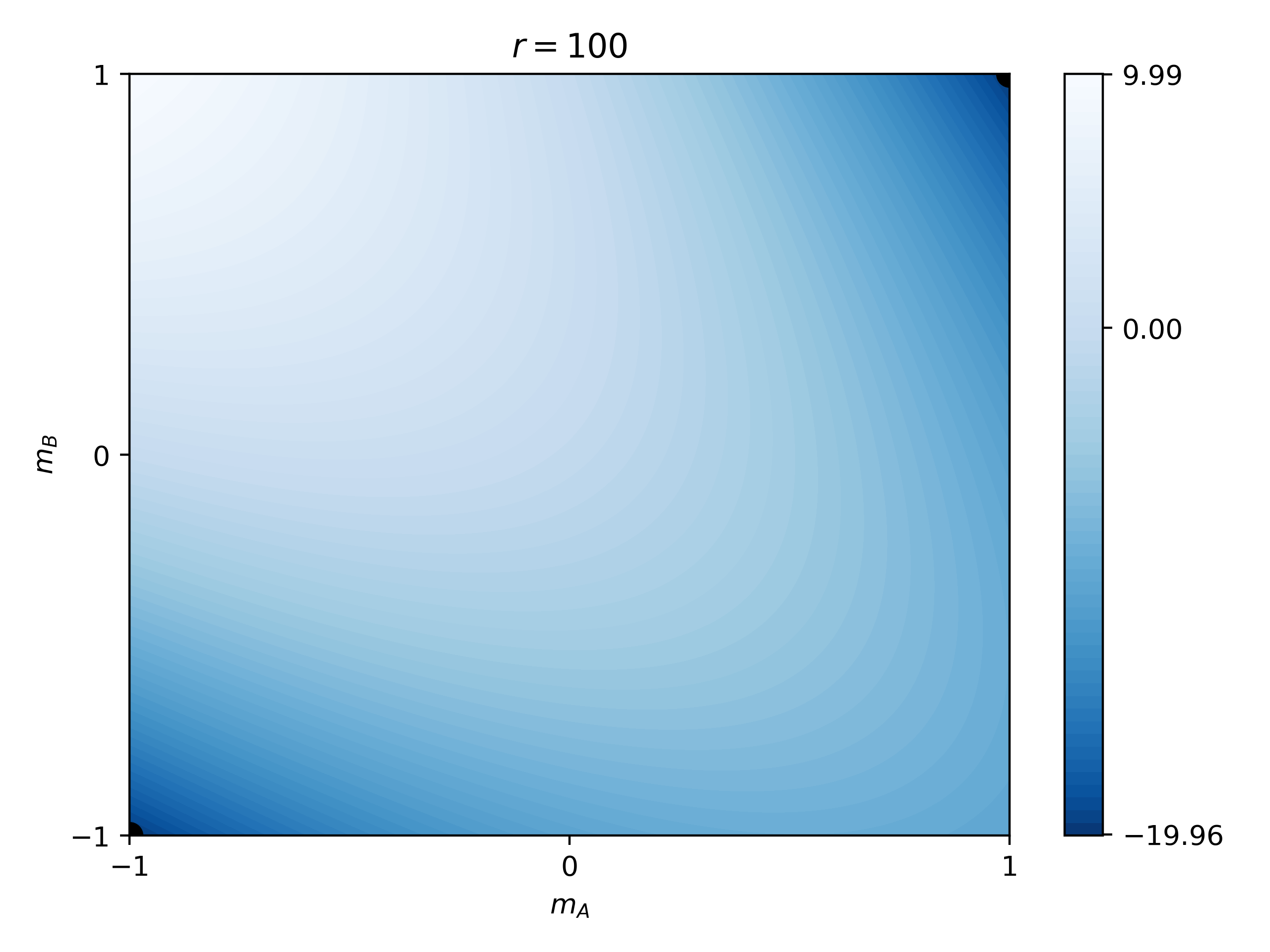}
  \caption{Mean-field free energy for different rationalities $r$. The parameters of the game are set to $\alpha=0.5, \; S=0.6$. In black, the local minima of the free energy, half-white and half-black are the saddle points. In the last figure, as $r>>1$ the minima stand almost at the corners $(1,1)$ and $(-1,-1)$, corresponding to full consensus.}
  \label{fig:free_en}
\end{figure}

\noindent
Connectivity plays a role of paramount relevance in a number of phenomena, including the dynamics of evolutionary games~\cite{santos02,moreno01}. Understanding the dynamics of a model whose entities are fully connected can be highly beneficial for assessing the effect of some more complex interaction topology. Therefore, before analysing the outcomes of the proposed model in $k$-regular networks, we observe those achieved by a fully-connected structure with a large number of nodes $N$.
In this setting, each agent has a degree $k=N-1\simeq N$. Also, a fraction $\alpha$ of agents prefer the $+1$ strategy (group A, $N_A=\alpha N$ in number), while the others prefer the strategy $0$ (group B, $N_B=N-N_A$).
Once the mapping (\ref{mapping_complete}), we are left with a bi-populated mean-field Ising model \cite{gallo2007bipartite,contucci2008phase,agliari2010two}, for which we derive the free energy in the large $N$ limit (\ref{free_en}), calculate the stationary points through the mean-field equations (\ref{mean_field_eq_games}) and predict the equilibrium states. \\
\\
The mapping for the complete network reads
\begin{equation}
\begin{split}
    &h_i  =  \frac{1-S}{4} N := h_A \quad \quad \quad \forall \;i\in A \\
    &h_j =  \frac{S-1}{4} N := h_B = -h_A  \quad \quad \forall \;j\in B \\
    &J = \frac{1+S}{4}          \\
    &\beta = R = \frac{r}{N}
\label{mapping_complete}
\end{split}
\end{equation}
\noindent Notice that we obtain a generalization of the Random Field Ising Model (RFIM)~\cite{imry1975random,galam1982new} with different numbers of sites with positive and negative fields. Considering group A and B's average magnetizations $m_{A/B}= \frac{1}{N_{A/B}}\sum_{i\in V_{A/B}}\sigma_i$ respectively (notice that $m_{A/B} = 2\rho_{A/B}-1$, where $\rho_{A/B}$ is the fraction of individuals of group $A/B$ playing $+1$), and the corresponding effective Hamiltonian
\begin{equation}
    \beta H = \frac{r}{N} \bigg[ -h_A \sum_{i\in V_A} \sigma_i - h_B \sum_{i\in V_B} \sigma_i -J\sum_{ij\in E}\sigma_i\sigma_j \bigg],
\end{equation}
the free energy (see \cite{gallo2007bipartite} for the derivation) reads
\begin{align}
    f(m_A,m_B; h_A,h_B,J,r,\alpha) = &-r \bigg[ \frac{J}{2} \bigg( \alpha^2 m_A^2 + (1-\alpha)^2 m_B^2 + 2\alpha(1-\alpha)m_A m_B \bigg) + \Tilde{h}_A\alpha m_A + \Tilde{h}_B(1-\alpha )m_B \bigg] + \nonumber \\
    & -\; \alpha I(m_A) - (1-\alpha) I(m_B)
\label{free_en}
\end{align}
\noindent where $\Tilde{h}_{A/B} := \frac{h_{A/B}}{<k>} = \frac{h_{A/B}}{N} = \pm \frac{1-S}{4}$ and $I(x) = -\frac{1+x}{2} \log(\frac{1+x}{2}) - \frac{1-x}{2} \log(\frac{1-x}{2})$ is the binary entropy function. 
The values of the magnetizations $m_A,m_B$ at equilibrium, indicated with $m_A^*,m_B^*$, correspond to the ones at the global minimum of the free energy functional. The stationary points of the latter functional are the solution(s) of the set of mean-field equations, obtained by setting to zero the derivatives of (\ref{free_en}) with respect to $m_A$ and $m_B$,

\begin{equation}
    \begin{cases}
        m_A = \tanh{\Bigg( r\bigg[  \Tilde{h}_A + J\bigg(\alpha m_A + (1-\alpha)m_B \bigg)    \bigg]\Bigg) } \\ 
        m_B = \tanh{\Bigg( r\bigg[  \Tilde{h}_B + J\bigg(\alpha m_A + (1-\alpha)m_B \bigg)    \bigg]\Bigg) }
    \end{cases}
\quad \quad
    \begin{cases}
        m_A = \tanh{\Bigg( \frac{r}{4}\bigg[ 1-S + (1+S)\bigg(\alpha m_A + (1-\alpha)m_B \bigg)    \bigg]\Bigg) } \\ 
        m_B = \tanh{\Bigg(  \frac{r}{4}\bigg[  -1 +S + (1+S)\bigg(\alpha m_A + (1-\alpha)m_B \bigg)    \bigg]\Bigg) }
    \end{cases}
\label{mean_field_eq_games}
\end{equation}
\noindent where, on the right, is the set of mean-field equations having substituted the Ising parameters with the ones of the game. In figure \ref{fig:free_en} is reported the free energy and its stationary points for four values of the rationality $r$, for $\alpha=0.5$. In this particular case, exploiting the symmetry, we can take $m^*_A+m^*_B$ as an order parameter and see that it undergoes a transition from being unique and zero up to a certain value of the rationality (inverse temperature), to show a positive and a negative value (two stable fixed points) after that point. Moreover, for very low rationalities ($r=0.1$), the only fixed point is localized around $m_A^* \simeq m_B^* \simeq 0$, meaning that the agents of both classes play $1$ or $0$ with approximately the same probability. For low rationalities ($r=2 $ in the figure), the fixed point is still unique but localized in the fourth quadrant ($m_A^*>0,m_B^*<0$), and a distance in average strategies between the classes emerges (polarization). For higher rationality ($r=4 $ in the figure), two fixed points corresponding to local minima appear as well as one saddle point: a phase transition of ferromagnetic type has occurred, and the system ends up in one or the other minimum with equal probability by spontaneous symmetry breaking. In the latter states, one class has managed to induce the other to play in the majority of its preferred strategy. \\
\\
In \cite{collet2014macroscopic}, it is found that at least for $h_A=h_B=0$ the equilibria of the Glauber (Logit) dynamics and their associated stability correspond to the stationary points of the free energy functionals, i.e. the solutions of the mean-field system (\ref{mean_field_eq_games}). We use the stationary states of the mean-field free energy to predict the magnetizations at equilibrium (global minimum) and, as a numerically tested approximation (see Figure \ref{fig: regular_simu and mf predictions}), to identify the relaxation points of the game dynamics (local minima), also for $\alpha \neq 0.5$.\\
\\
Last, we mention that the best-response regime is recovered for $r\rightarrow \infty$: in this regime (see \cite{galam1997rational} and, using a dynamical approach, appendix \ref{sec:best_response}), the fully polarized state $(m_A=1,m_B=-1)$ is a stable fixed point of the dynamics only for $S< \frac{\alpha}{1-\alpha}$ (assuming without loss of generality that $\alpha\leq 0.5$, see \eqref{mf threshold best response}), otherwise the only possible equilibria are the two full consensus states $(1,1)$ and $(-1,-1)$.

\subsection{Random k-regular networks}
\begin{figure}    
    \begin{minipage}[b]{0.3\textwidth}

        \includegraphics[width=\linewidth]{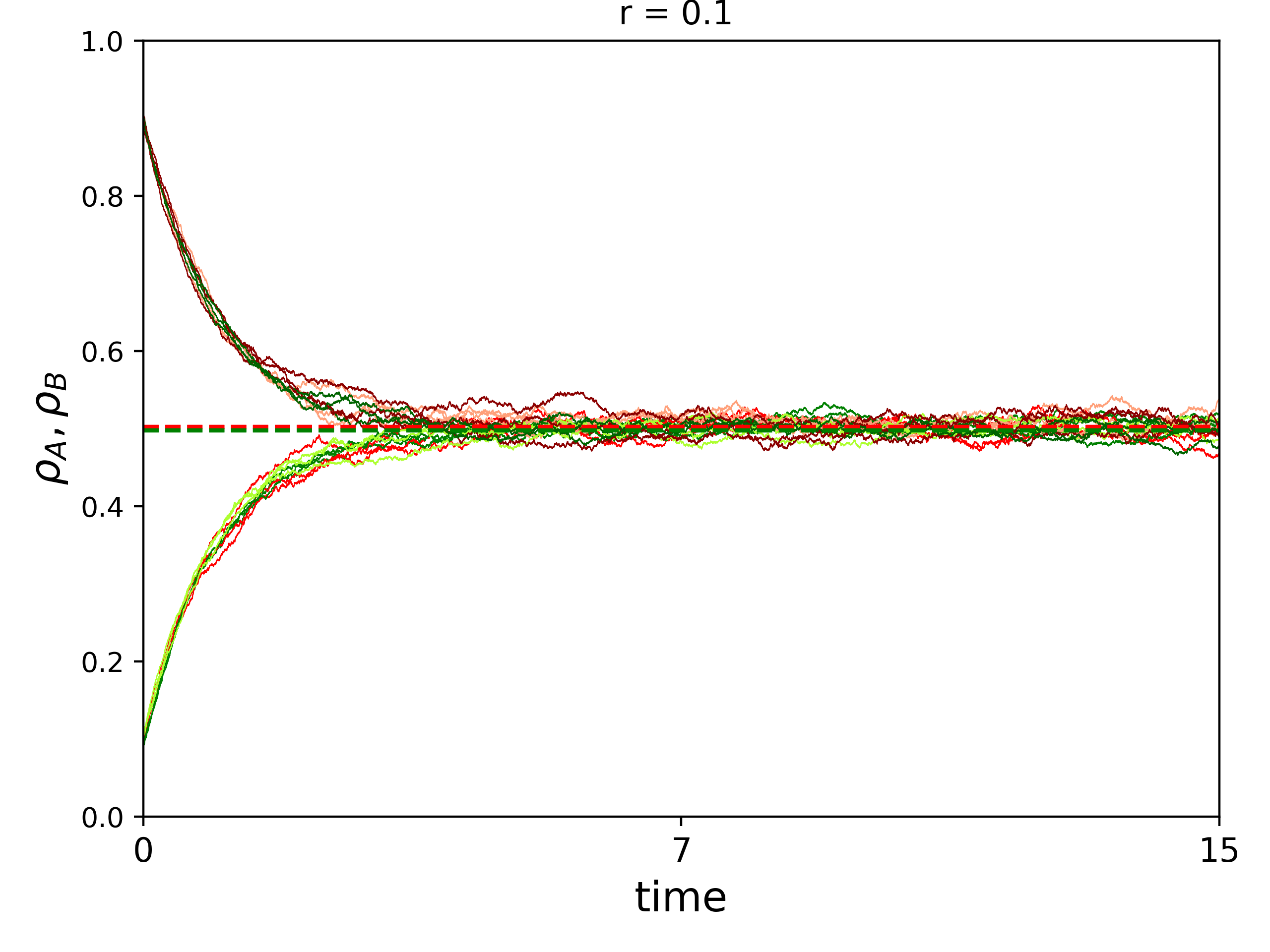}
    \end{minipage}
    \hfill
    \begin{minipage}[b]{0.3\textwidth}
        \includegraphics[width=\linewidth]{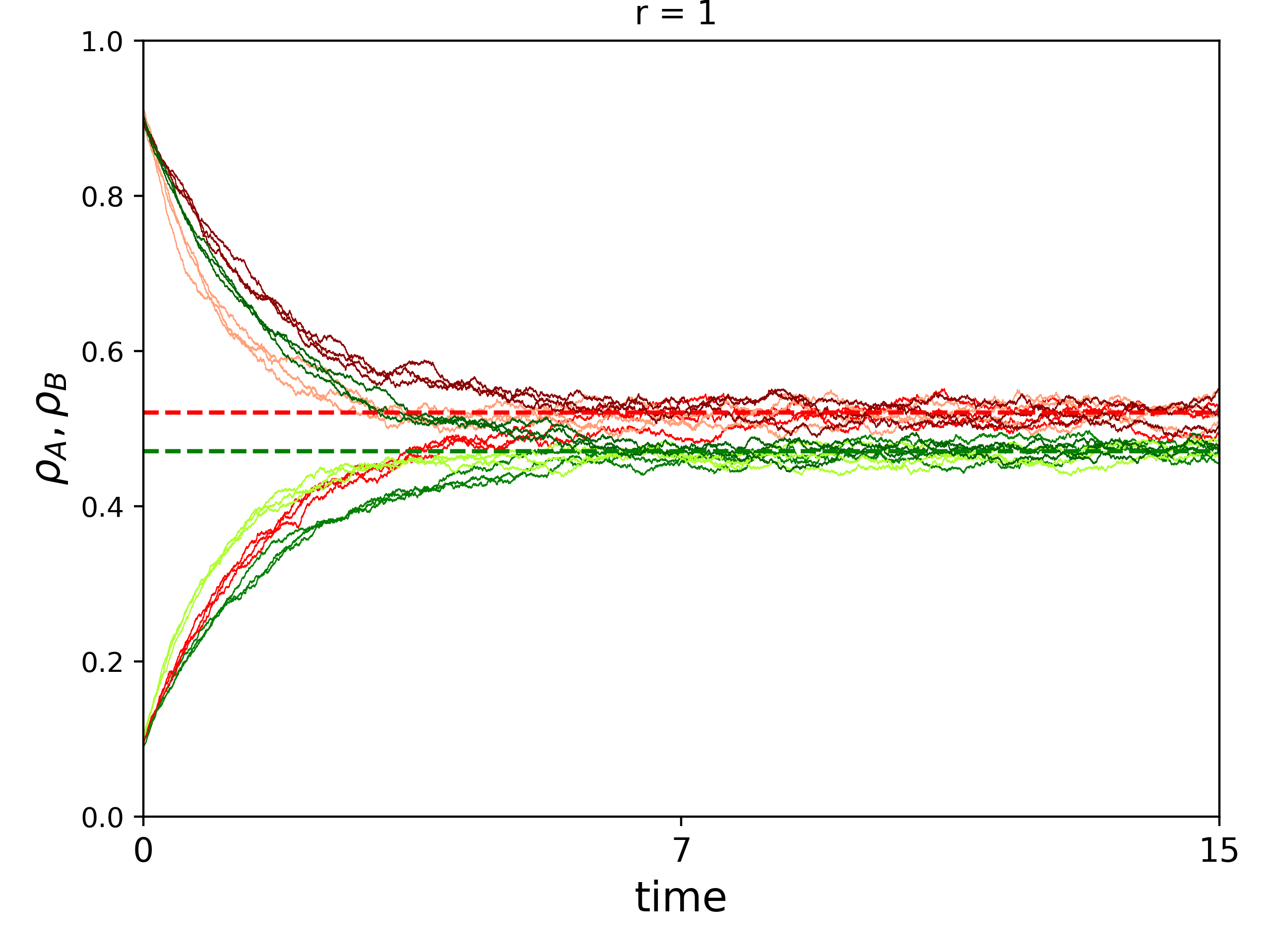}
    \end{minipage}
    \hfill
    \begin{minipage}[b]{0.3\textwidth}
        \includegraphics[width=\linewidth]{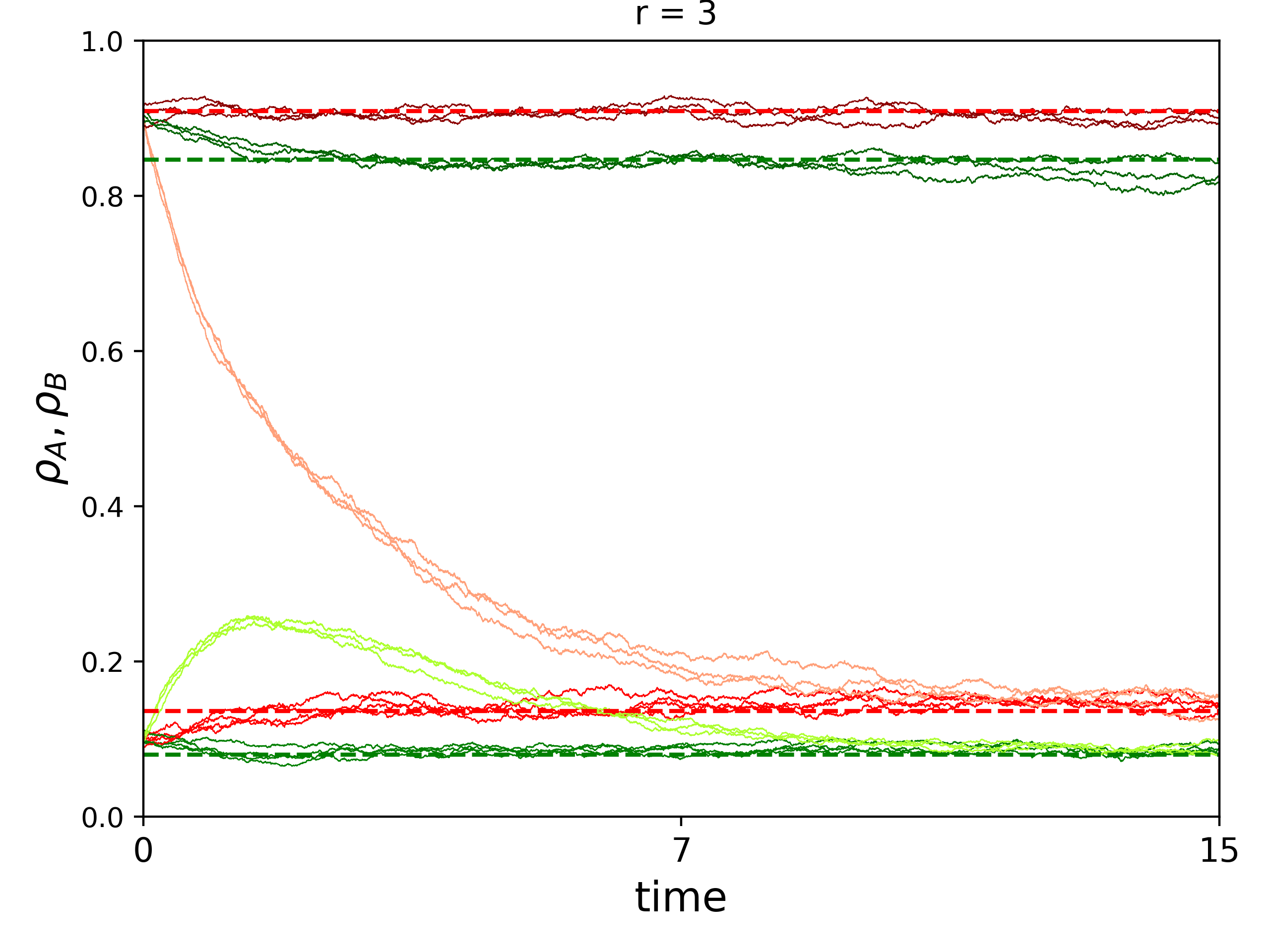}
    \end{minipage}
    
    \vspace{0.5cm} % Adjust the vertical spacing between the rows
    
    \begin{minipage}[b]{0.3\textwidth}
        \includegraphics[width=\linewidth]{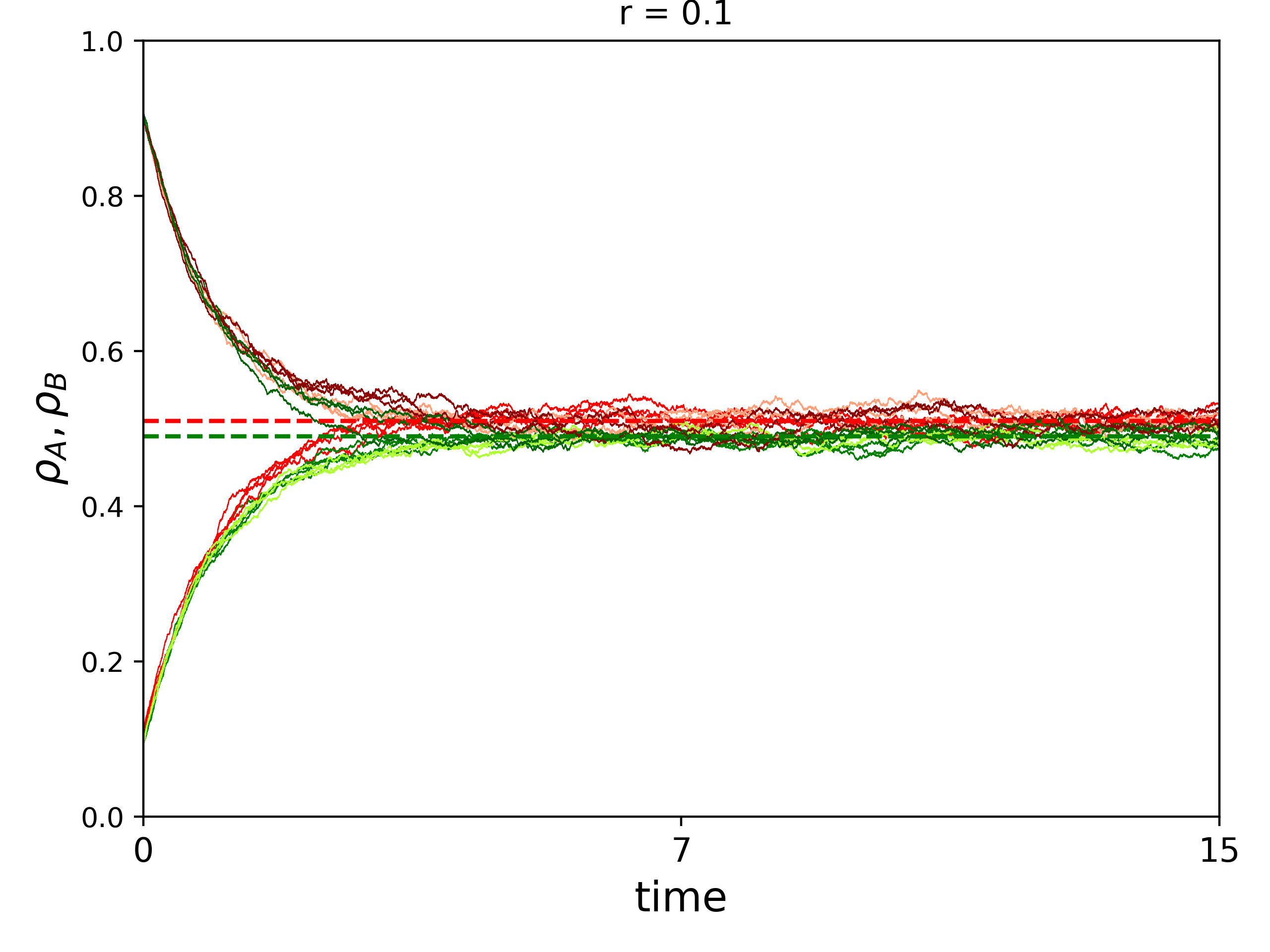}
    \end{minipage}
    \hfill
    \begin{minipage}[b]{0.3\textwidth}
        \includegraphics[width=\linewidth]{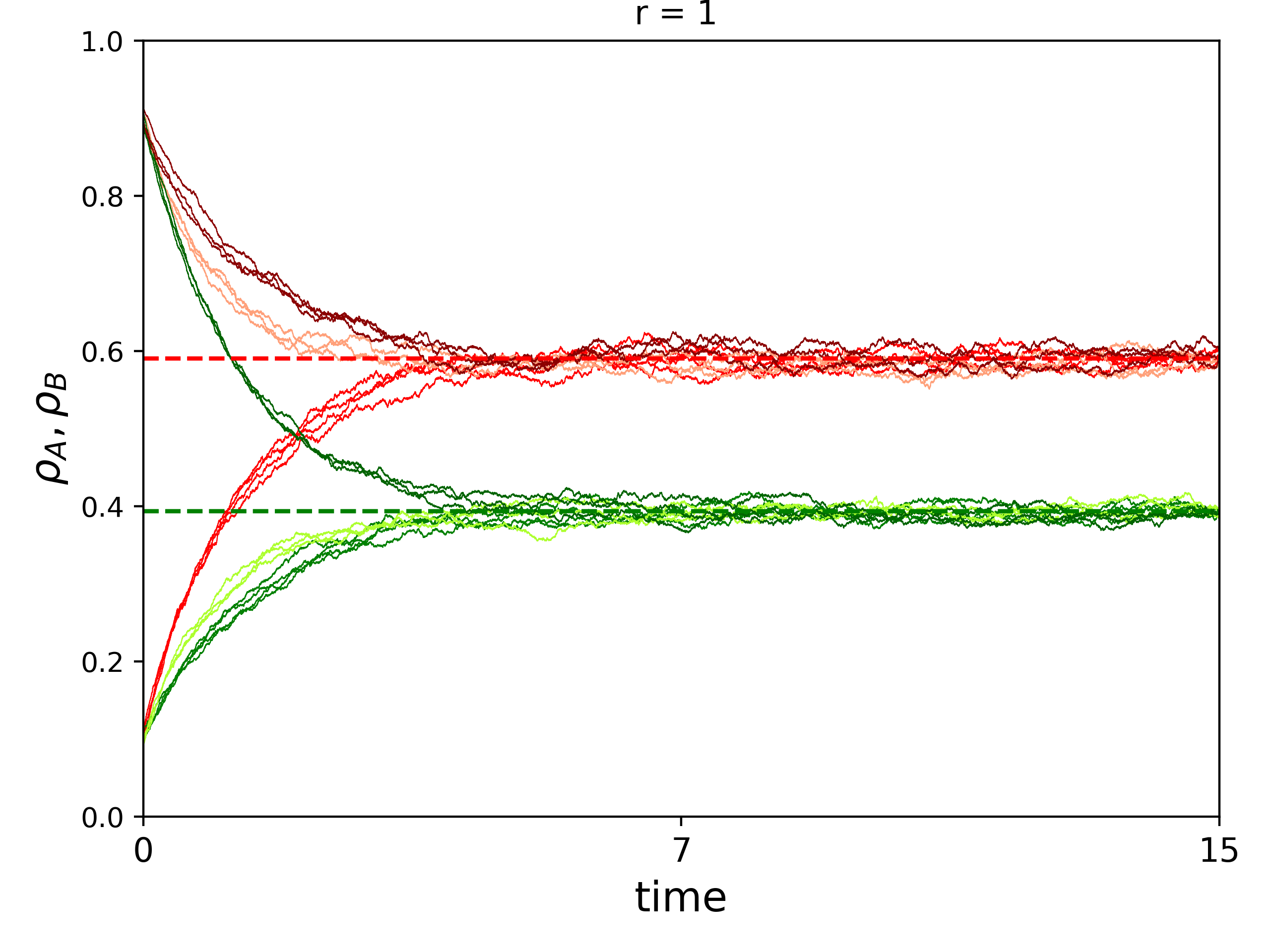}
    \end{minipage}
    \hfill
    \begin{minipage}[b]{0.3\textwidth}
        \includegraphics[width=\linewidth]{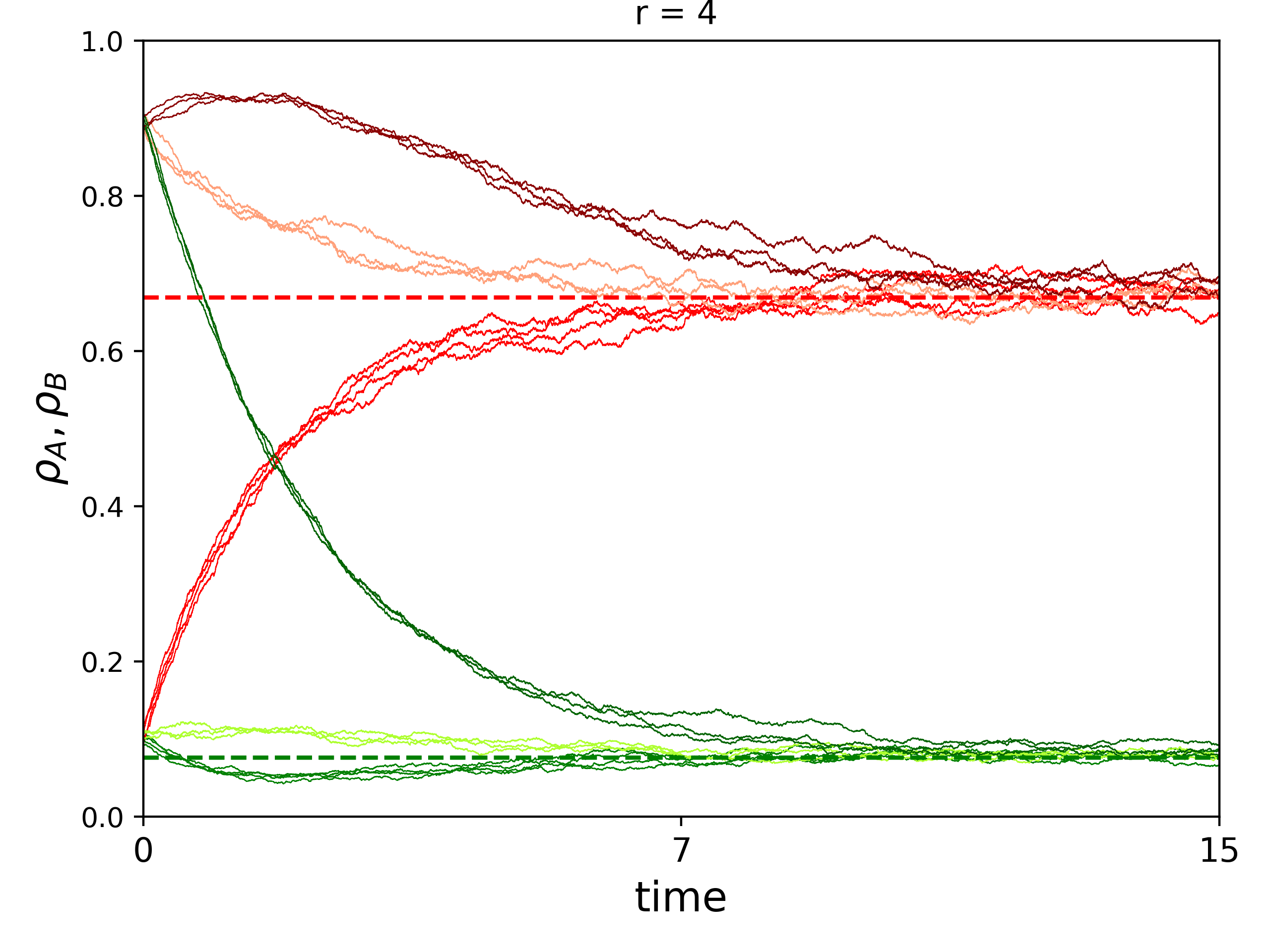}
    \end{minipage}

    \caption{\textbf{Mean-field predictions and simulations.} The network is $k$-regular with $N=5000$ agents and $k=30$. The fraction of agents with $+1$ preference (group $A$) is $\alpha=0.4$. The figures show both the mean-field predictions for the stationary density of $+1$ spins with preference $+1$ ($\rho_A^*$) of $+1$ spins with preference $0$ ($\rho_B^*$), and the behaviour of $\rho_A(t),\rho_B(t)$ as a function of time for 10 simulations of the game with different initial conditions $(\rho_A(0),\rho_B(0))$, differentiated according to the couples of colors: light green/light blue correspond to $(0.9,0.1)$, green/red to $(0.1,0.1)$, dark green/dark red to $(0.9,0.9)$. The dashed red/green horizontal lines are the mean-field predictions, i.e., the solutions of the mean-field system corresponding to the local minima of the mean-field free energy, respectively  $\rho_A^*= \frac{1+m_A^*}{2}$ and $\rho_B^*= \frac{1+m_B^*}{2}$. In the upper row, $S=0.8$, while in the lower one $S=0.2$. The rationalities $r$ are specified in the titles of the plots.}

    \label{fig: regular_simu and mf predictions}
\end{figure}

\noindent
Now, let us consider $k$-regular random networks~\cite{van2009random}, i.e. networks whose nodes have the same finite degree $k$ with connections drawn uniformly randomly. 
In this setting, by parametrizing $R=\frac{r}{k}$, the mean-field equation system coincides with ~\eqref{mean_field_eq_games}.
That is due to the linear dependency of the field $h_i$ on its degree $k_i$ (eq. \eqref{mapping general}) and the homogeneity of degrees.
Nevertheless, the mean-field approach is exact in a complete network for $N\rightarrow \infty$, while the same does not apply to regular networks. We expect the approximation to work better in denser networks, i.e., networks having a higher $k$. 
In figure \ref{fig: regular_simu and mf predictions}, we show the mean-field predictions for the relaxation state (local minima of the mean-field free energy) by comparing the simulation outcomes of multiple games:  we set $\alpha= 0.4$, thus $S^*=0.67$ (eq. \ref{mf threshold best response}), and vary the rationality through $r$ for $S=  0.8 >S^*$ (upper plots) and for $S=  0.2<S^*$ (lower plots). The effects of the finite degrees are discussed, at least in the regime of infinite rationality, in the appendix \ref{sec:best_response}. \\
\\

\subsection{Polarization and Rationality}
\begin{figure}[h!]

    \includegraphics[width=0.6\linewidth]{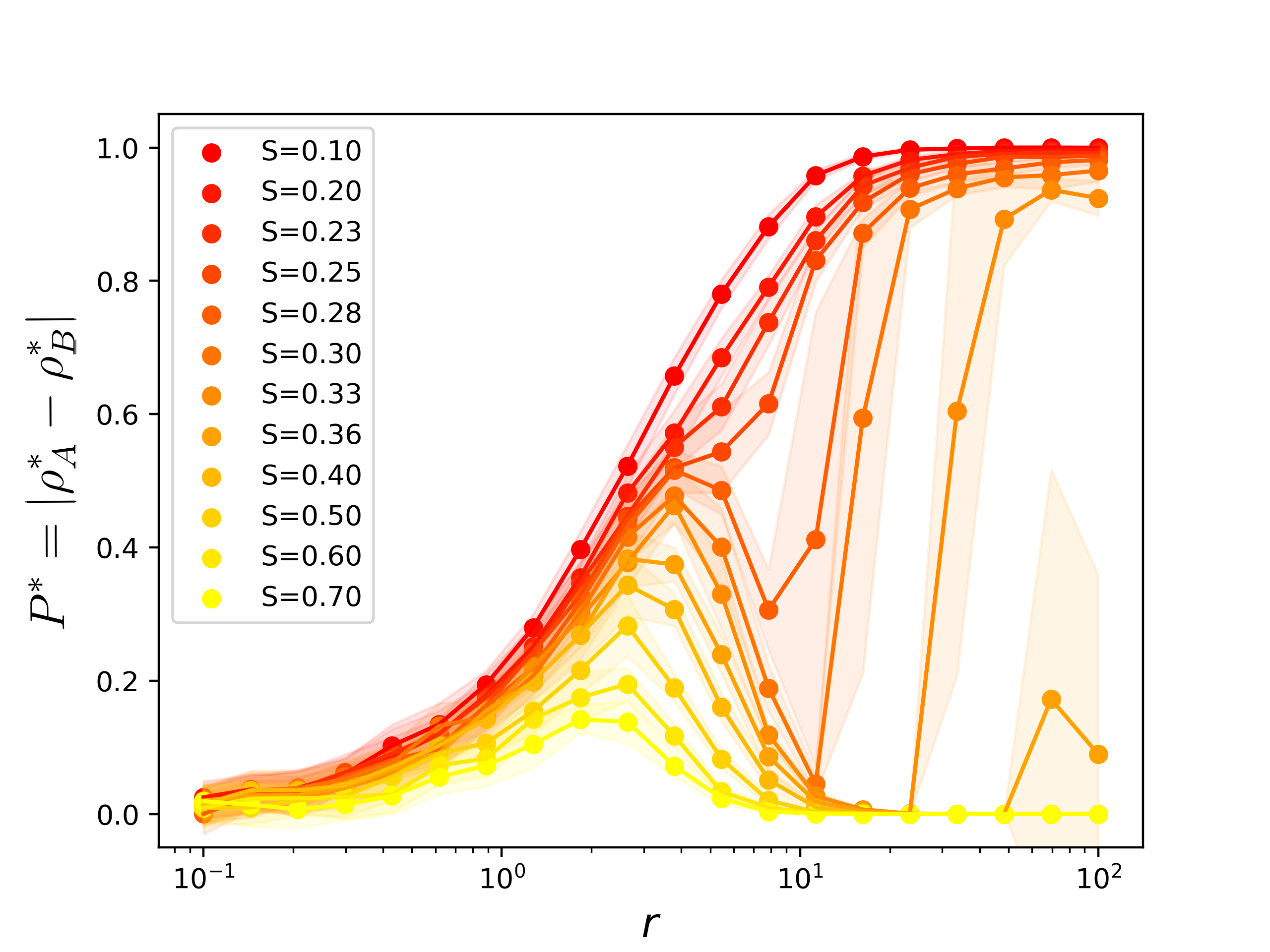}

    \caption{\textbf{Polarization at the stationary state, as a function of the rationality.} We consider simulated games on a regular graph of $N=1000$ agents with degree $k=30$ and $\alpha= 0.4$, for various intensities of the preference $S$ as in the legend. The initial condition corresponds to everyone choosing his preferred strategy. The dots correspond to the polarization value at the state after relaxation and are averaged over 10 independent realizations of the model.}
    \label{fig: polarization}
\end{figure}

\noindent
The mapping between the BoS game (on networks) and the Ising model allowed us to gain a preliminary overview of the effects of bounded rationality in these dynamics by an analytical approach. Now, we study the relationships between group polarisation and bounded rationality by numerical simulations. 
For clarity, indicating with $\rho_{A/B}(t)$ the evolution of the density of the agent with $+1$ opinion in the two groups, the polarization is defined as the distance between the average opinions of the two groups, i.e. $P(t)=|\rho_A(t)-\rho_B(t)|$. Simulations implemented on a regular graph composed of $N=1000$ agents, with a finite degree $k=30$, $\alpha= 0.4$, consider various intensities of the preference $S$ --- see the legend of figure~\ref{fig: polarization}. 
In all cases, in the beginning, the agents choose their preferred strategy, thus $\rho_A(0)=1, \rho_B(0)=0$. In the figure, the dots are the values of the polarization at the stationary state $P^*=|\rho_A^*-\rho_B^*|$ for different rationalities $r$ ($R=\frac{r}{k}$), where $\rho_{A/B}^*$ are the averages of the densities of agents playing $1$ at the state reached after relaxation.\\
We see that the behaviour of the polarization as a function of rationality is highly non-trivial. For all the values of $S$, at low rationalities, the polarization shows a monotonic behaviour for low $r$. Then, polarization can either keep increasing (high $S$) until it becomes maximum at infinite rationality (full polarization) or else it can decrease after reaching a peak (middle and low $S$). For middle $S$, at a point, it suddenly bumps up to reaching almost full polarization while, for low $S$, it stays very close to zero (almost consensus) even for infinite rationality. 
This behaviour roughly follows the mean-field predictions but with some deviations that, in our interpretation, are due to the finiteness of the degrees and the fluctuations induced by limited rationality. The former may generate a cascade effect leading the system to the state corresponding to consensus at the majority's preferred opinion (see section \ref{sec:best_response} in the appendix for an extended analysis), while limited rationality increases the fluctuations and thus the possibility to fall into the basin of attraction of the consensus points. Both the effects favour the approach to the (almost-)consensus states even for stronger preference intensities (low $S$) concerning what is predicted within the mean-field approximation (see Fig. \ref{fig best response}d ).

\subsection{Asymmetric games of Hernandez et al.} \label{subsec:hernandez} 
\noindent Both the models of Broere et al. \cite{broere2017network} and Hernandez et al. \cite{hernandez2013heterogeneous} consider two classes of agents, $A,B$, with conflicting preferences of equal intensities. The difference is in the fact that Hernandez et al. consider a further reward for the agent's choice independent of the neighbours' choices, i.e. what we call single term $\chi^{(1)}$ in \eqref{general_payoff_eq}, which is greater if the chosen strategy corresponds to the preferred one. Thus, for the model of Hernandez et al., with rewards $0<\beta<\alpha<2\beta$ (see the payoff matrix of \cite{hernandez2013heterogeneous}), we have in our notation
\begin{align}
    &r_A^{(1)} = \alpha    \quad \quad r_A^{(0)} = \beta &r_B^{(1)} = \beta   \quad \quad r_B^{(0)} = \alpha \nonumber \\ 
    &a_A^{(11)} = \alpha    \quad \quad a_A^{(10)}=0   &a_B^{(11)} = \beta    \quad \quad a_B^{(10)}=0 \nonumber\\
    &a_A^{(01)} = 0     \quad \quad a_A^{(00)}=\beta  &a_B^{(01)} = 0    \quad \quad a_B^{(00)}=\alpha 
\end{align}
Thus, by following the mapping towards the Ising model reported in the appendix, one notes that $\chi^{(1)}$ just adds a term in the magnetic field (eq. \ref{hernandez_mapping_h}), which is subleading in $k$.

\section{Discussion}\label{sec:discussion}
\noindent
In this work, we studied asymmetric games with bounded rationality, mapping their dynamics to that of an Ising model. 
We consider agents endowed with a fixed preference in a binary opinion system, leading to the formation of two communities (or groups).  
Namely, each group is biased towards one of the two opinions, say $0$ and $1$, respectively. Also, while preferences cannot change, i.e. are fixed, agents can change opinions. 
The agents' rationality, corresponding to the system temperature~\cite{javarone03}, affects the population dynamics~\cite{javarone05}. Then, mapping the agent population to the Ising model, the convergence of the agents towards stable opinions resembles a phase transition. 
For instance, low rationality entails the existence of a single equilibrium. Conversely, higher values of rationality may lead the system towards configurations with two or more stable states.
In this context, an equilibrium corresponds to a state with two groups having an average density stable over time.
While the mathematical formulation of the model is independent of the agent population structure, we performed the analysis considering a complete and a $k$-regular network. The analytical calculations rely on the mean-field approximation that, as known, is not exact for $k$-regular networks. Therefore, for the second structure (i.e. the $k$-regular network), we investigated the effects of finite degrees using numerical simulations. Our analyses include two conditions, i.e. bounded and infinite rationality. \\
It turns out that bounded rationality and preference intensity determine a series of possible scenarios characterized by different levels of polarization (i.e. the distance between the groups in terms of average opinion). Within the mean-field theory, we find the following cases:
\begin{itemize}
    \item low rationality: there exists a unique stable state weakly polarized (whose polarization depends on the preference intensity);
    \item high rationality and low preference intensity: there exist two fixed points corresponding to almost-consensus states at one and the other opinion (so with very low polarization);
    \item high rationality and high preference intensity: there exists instead a single fixed point corresponding to a strongly polarized state;
    \item very high rationality and in the limit of infinite rationality, for high preference intensities: two consensus fixed points pop up and stand together with the polarized one.
\end{itemize}
In the presence of multiple fixed points, the reached one depends on the initial conditions and single realizations.
The finiteness of the network and the network's degrees, breaking the mean-field assumptions, favours in general (almost-)consensus states, especially for low rationality.\\
Interpreting the rationality as the agents' attention to minimize their personal and social dissonances (as in \cite{dalege2023networks,jarema2022private}), increasing the latter from a weakly polarized state may lead the system to an almost-complete consensus state or a very polarized one, depending on the preference intensity.
Accordingly, a policy-maker aiming to curb conflicts in a community should be aware of groups of individuals with fixed conflicting preferences towards specific issues ---see also~\cite{javarone04} on this topic.
For instance, raising attention to one of these issues can divide public opinion and severely increase polarization. 
On the other hand, if the preference intensities are sufficiently low, raising attention to an issue makes public opinion reach a low conflict, almost-consensus state, albeit some individuals have falsified their preferences.\\
Beyond that, our work sheds light on the dynamics of asymmetric games, typically studied by numerical simulations and infinite rationality~\cite{galam1997rational}. 
Let us remark that the mapping to the Ising model is performed by assuming opposite preferences of equal intensity (Broere's model), Logit-rule as a dynamical rule for the opinion update, and homogeneous rationality in the population. 

\section{Conclusion}\label{sec:conclusion}
\noindent
In summary, our work clarifies some aspects of the dynamics of asymmetric games through random field models. Interestingly, the achieved results find interpretations within the context of opinion dynamics and the formation of coordination.
While previous investigations, oriented towards social systems, attempted to describe similar scenarios, in this investigation, we focus on asymmetric games played by agents endowed with bounded rationality and map them to random field Ising models.
As reported above, results show the emergence of opinion polarization and consensus, depending on the agents' rationality.
In light of that, we deem our study sheds light on relevant aspects of asymmetric games, and the proposed formalisation in terms of random fields can support further works in this direction.
Notwithstanding, several aspects still deserve attention. To cite a few, future investigations may study the role of heterogeneous networks (e.g. small-world structures and scale-free networks \cite{barabasi2013network}, multi-layer networks \cite{boccaletti2014structure}, and networks with higher-order interactions \cite{civilini2023explosive}), the effect of homophily reflecting the preferences' assignment, as in a previous work \cite{zimmaro2023voter}, and that of larger opinion systems.
Eventually, further developments can relate to the framework of evolutionary game theory. For instance, treating opinions as strategies, the preferences could describe the agents' attitude towards cooperation or defection in dilemma games. 
To conclude, we remark that our results rely on the mapping to the Ising model. Thus, further developments in applying statistical mechanics to opinion dynamics and (evolutionary) game theory~\cite{javarone00} can exploit the formalism we proposed in this work.

\appendix
\section{Asymmetric network games and Ising models} \label{sec:general_mapping} 
\noindent
A general coordination or anticoordination game on a network $G=(V,E)$ with arbitrary order interactions, $i=1,...,N$ agents, two strategies $x_i=\{0,1\}$ and response (dynamical rule) depending only on the agent's possible payoffs, can be expressed through the payoff functions of each agent
$$\pi_i(x_i,\mathbf{x_{-i}}) = \chi_i^{(1)}(x_i) + \chi_i^{(2C)}(x_i)\sum_{j\in\partial i} I_{\{x_i=x_j\}} + \chi_i^{(2AC)}(x_i)\sum_{j\in\partial i} I_{\{x_i\neq x_j\}} + \chi_i^{(3C)}(x_i)\sum_{<ijk>\in G} I_{\{x_i=x_j=x_k\}} + ...$$
where $\chi_i^{(1)}(x_i)$ is the reward for agent $i$ choosing strategy $x_i$, $\chi_i^{(2C)}(x_i)$ the reward for each successfull coordination at $x_i$, $\chi_i^{(2AC)}(x_i)$ for each pairwise anti-coordination, $\chi_i^{(3C)}(x_i)$ for each coordination in an hyperedge with two neighbours of $i$ and so on. \\
We consider so far only single and pairwise interactions, so the payoff before reduces to 
\begin{equation}
    \pi_i(x_i,\mathbf{x_{-i}}) = \chi_i^{(1)}(x_i) + \chi_i^{(2C)}(x_i)\sum_{j\in\partial i} I_{\{x_i=x_j\}} + \chi_i^{(2AC)}(x_i)\sum_{j\in\partial i} I_{\{x_i\neq x_j\}}
\label{general_payoff_eq}
\end{equation}
Now we can use the matrix representation of the payoff, expliciting
\begin{equation}
    \chi_i^{(1)} = 
    \begin{cases}
        r_i^{(1)}\;\;\;\;\mbox{if}\;\;\;x_i=1 \\
        r_i^{(0)}\;\;\;\;\mbox{if}\;\;\;x_i=0 \\
    \end{cases}
\end{equation}
\begin{equation}
    \chi_i^{(2C)} = 
    \begin{cases}
        a_i^{(11)}\;\;\;\;\mbox{if}\;\;\;x_i=1 \\
        a_i^{(00)}\;\;\;\;\mbox{if}\;\;\;x_i=0 \\
    \end{cases}
\end{equation}
\begin{equation}
    \chi_i^{(2AC)} = 
    \begin{cases}
        a_i^{(10)}\;\;\;\;\mbox{if}\;\;\;x_i=1 \\
        a_i^{(01)}\;\;\;\;\mbox{if}\;\;\;x_i=0 \\
    \end{cases}
\end{equation}
considering for the moment $r_i^{(1)} = r_i^{(0)} = 0$, for each pairwise interaction of the agent $i$
\begin{equation}
\begin{array}{cc|c|c|}
      & \multicolumn{1}{c}{} & \multicolumn{2}{c}{\mathbf{\cdot}}\\
      & \multicolumn{1}{c}{} & \multicolumn{1}{c}{1}  & \multicolumn{1}{c}{0} \\\cline{3-4}
      \multirow{2}{*}{$\mathbf{i}$}  & 1 & (a^{(11)}_i,\cdot) & (a^{(10)}_i,\cdot) \\ \cline{3-4}
      & 0 & (a^{(01)}_i,\cdot) & (a^{(00)}_i,\cdot) \\\cline{3-4}
\end{array}
\end{equation}
Now using the vector representation $\vec{a} = (a^{(11)},a^{(10)},a^{(01)},a^{(00)})$, we decompose such vector in four orthogonal components
\begin{align}
    &\vec{\alpha} = (1,1,-1,-1)\\
    &\vec{\beta} = (1,-1,1,-1)\\
    &\vec{\gamma} = (1,-1,-1,1)\\
    &\vec{\eta} = (1,1,1,1)\
\end{align}
so that a generic payoff matrix can be written in the so called normal-form as
\begin{equation}
\begin{array}{cc|c|c|}
      & \multicolumn{1}{c}{} & \multicolumn{2}{c}{\mathbf{\cdot}}\\
      & \multicolumn{1}{c}{} & \multicolumn{1}{c}{1}  & \multicolumn{1}{c}{0} \\\cline{3-4}
      \multirow{2}{*}{$\mathbf{i}$}  & 1 & (\alpha_i+\beta_i+\gamma_i+\eta_i,\cdot) & (\alpha_i-\beta_i-\gamma_i+\eta_i,\cdot) \\ \cline{3-4}
      & 0 & (-\alpha_i+\beta_i-\gamma_i+\eta_i,\cdot) & (-\alpha_i-\beta_i+\gamma_i+\eta_i,\cdot) \\\cline{3-4}
\end{array}
\end{equation}
from which we interpret $\alpha_i$ as the reward to choose the preferred action, $\gamma_i$ the reward for achieving coordination, $\beta_i$ the external payoff depending only on the other agent's action and $\eta_i$ as a free independent reward, all for agent $i$. The coefficients come from the system $\vec{a_i} = \alpha_i \vec{\alpha} + \beta_i \vec{\beta} + \gamma_i \vec{\gamma} + \eta_i \vec{\eta}$, i.e.,
\begin{equation}
    \begin{bmatrix}
    1 & 1 & 1 & 1 \\
    1 & -1 & -1 & 1 \\
    -1 & 1 & -1 & 1 \\
    -1 & -1 & 1 & 1
    \end{bmatrix}
    \begin{bmatrix}
    \alpha  \\
    \beta   \\
    \gamma  \\
    \eta 
    \end{bmatrix}
    =
    \begin{bmatrix}
    a^{(11)}\\
    a^{(10)} \\
    a^{(01)} \\
    a^{(00)} 
    \end{bmatrix}
\end{equation}
which inverted gives 
\begin{equation}
\begin{split}
    &\alpha_i = \frac{a_i^{(11)}+a_i^{(10)} - a_i^{(01)} - a_i^{(00)}}{4}\quad \quad \quad  \beta_i = \frac{a_i^{(11)}-a_i^{(10)} + a_i^{(01)} - a_i^{(00)}}{4}\\ &\gamma_i =\frac{a_i^{(11)}-a_i^{(10)} - a_i^{(01)} + a_i^{(00)}}{4} \quad  \quad \quad \eta_i = \frac{a_i^{(11)}+a_i^{(10)} + a_i^{(01)} + a_i^{(00)}}{4}
\end{split}
\end{equation}
\\
When the response function determining the dynamics depends only on the difference between the payoffs for the selected agent associated to the two strategies (e.g. best response, stochastic best response, Logit-rule etc.), we have that 
\begin{align}
    \pi(1,\mathbf{x_{-i}}) - \pi(0,\mathbf{x_{-i}})& =  \bigg[ a^{(10)}_i \sum_{j\in\partial i}(1-x_j) + a^{(11)}_i \sum_{j\in\partial i}x_j\bigg] - \bigg[ a^{(00)}_i \sum_{j\in\partial i}(1-x_j) + a^{(01)}_i \sum_{j\in\partial i}x_j\bigg]  = \\
    & = (a^{(11)}_i - a^{(01)}_i ) \sum_{j\in\partial i}x_j -  ( a^{(00)}_i - a^{(10)}_i ) \sum_{j\in\partial i}(1-x_j) =\\
    & = 2(\alpha_i+\gamma_i) \sum_{j\in\partial i}x_j -  2( -\alpha_i +\gamma_i) \sum_{j\in\partial i}(1-x_j)
\end{align}
having applied the decomposition before. \\
As the dynamics depends only on the preference and coordination terms, without loss of generality we can simplify the game considering the payoff matrix without the $\beta$ and $\eta$ terms
\begin{equation}
\begin{array}{cc|c|c|}
      & \multicolumn{1}{c}{} & \multicolumn{2}{c}{\mathbf{\cdot}}\\
      & \multicolumn{1}{c}{} & \multicolumn{1}{c}{1}  & \multicolumn{1}{c}{0} \\\cline{3-4}
      \multirow{2}{*}{$\mathbf{i}$}  & 1 & (\alpha_i+\gamma_i,\cdot) & (\alpha_i-\gamma_i,\cdot) \\ \cline{3-4}
      & 0 & (-\alpha_i-\gamma_i,\cdot) & (-\alpha_i+\gamma_i,\cdot) \\\cline{3-4}
\end{array}
\end{equation}
Moreover \cite{guisasola2020potential}, such game has a potential if and only if 
\begin{equation}
    \gamma_i = \gamma_j \quad \quad \forall \;ij\in E
\end{equation}
so, assuming that the network has a single connected component, if and only if 
\begin{equation}
    (a_i^{(11)}+a_i^{(00)}) - (a_i^{(01)}+a_i^{(10)}) = C\quad \quad \forall\; i=1,...,N
\end{equation}
with the same constant $C$. Thus, the game evolving with Logit rule is mappable to an Ising model evolving with Glauber dynamics \cite{szabo2007evolutionary, szabo2016evolutionary}. \\
Considering $\gamma_i=\gamma \;\;\forall i$ and manipulating the expression of the payoff difference we get
\begin{equation}
\begin{split}
    \pi(1,\mathbf{x_{-i}}) - \pi(0,\mathbf{x_{-i}})& = 2\sum_{j\in\partial i} (\alpha_i+ \gamma(2x_j-1)) =\\
    & = 2k_i\alpha_i + 2\gamma\sum_{j\in\partial i}(2x_j-1)
\end{split}
\end{equation}
where $k_i$ is the degree of node $i$.\\
We then pass to the spin variables $\boldsymbol{\sigma}=\{-1,+1\}^N$ through $\sigma_i = 2x_i-1\;\;\;\forall\;i=1,...N$, so
\begin{equation}   
    \pi(1,\mathbf{x_{-i}}) - \pi(0,\mathbf{x_{-i}}) = \pi(1,\mathbf{\sigma_{-i}}) - \pi(-1,\mathbf{\sigma_{-i}}) = 2k_i\alpha_i + 2\gamma\sum_{j\in\partial i}\sigma_j
\end{equation}\
Successively, we write the energy difference between two configurations after the flipping of a spin of an Ising model with fields $\boldsymbol{h}=\{h_j\}_{j\in V}$ and homogeneous couplings $J_{ij}=J\;\forall\;\;ij\in E$, on the same network
\begin{equation}
    H(\sigma_i=1,\mathbf{\sigma_{-i}}) - H(\sigma_i=-1,\mathbf{\sigma_{-i}}) =  -2h_i - 2J\sum_{j\in\partial i}\sigma_j
\end{equation}
From that we get immediately the mapping of the general game on network evolving with Logit rule with rationality $R$ to an Ising model on the same network evolving with Glauber dynamics with Hamiltonian
\begin{equation}
    H = -\sum_i h_i\sigma_i - J\sum_{ij\in E}\sigma_i\sigma_j
\end{equation}
and inverse temperature $\beta$, respectively 
\begin{equation}
      P_i(x_i) = \frac{e^{R[\pi_i(x_i;\theta_i) - \pi_i(0;\theta_i)]} }{1+e^{R[\pi_i(1;\theta_i) - \pi_i(0;\theta_i)]}} 
      \quad \quad \quad 
      P_i(\sigma_i) = \frac{e^{-\beta[H(\sigma_i,\mathbf{\sigma_{-i}}) - H(-1,\mathbf{\sigma_{-i}})]}}{1+e^{-\beta[H(+1,\mathbf{\sigma_{-i}}) - H(-1,\mathbf{\sigma_{-i}})]}}
\end{equation}
The mapping reads
\begin{align}
    &h_i = k_i\alpha_i =  \frac{(a_i^{(11)}+a_i^{(10)}) - (a_i^{(00)}+a_i^{(01)})}{4} k_i \;\;\;\;\;\;\;\;\;
    \;\;\;\;\;\;\;\;\;i=1,...,N\\
    &J = \gamma = \frac{(a_i^{(11)}+a_i^{(00)}) - (a_i^{(01)}+a_i^{(10)})}{4} \\
    &\beta = R
\end{align}
Last, it is easy to see that the single term $\chi_i^{(1)}$ adds a term to the payoff difference of $r^{(1)}_i - r_i^{(0)}$, so modifies the Ising magnetic fields in the way 
\begin{align}
    &h_i = k_i\alpha_i + \frac{r^{(1)}_i - r_i^{(0)}}{2}  =   \frac{(a_i^{(11)}+a_i^{(10)}) - (a_i^{(00)}+a_i^{(01)})}{4} k_i + \frac{r^{(1)}_i - r_i^{(0)}}{2}  \;\;\; \;\;\;\;\;\;\;\;\;i=1,...,N \label{hernandez_mapping_h}\\
    &J = \gamma = \frac{(a_i^{(11)}+a_i^{(00)}) - (a_i^{(01)}+a_i^{(10)})}{4} 
\end{align}

\section{Mean-field results for best-response} \label{sec:best_response} 
\begin{figure}[htbp]
  \centering
  \subfigure[$\;k = 300,\; S=0.5$]{\includegraphics[width=0.42\textwidth]{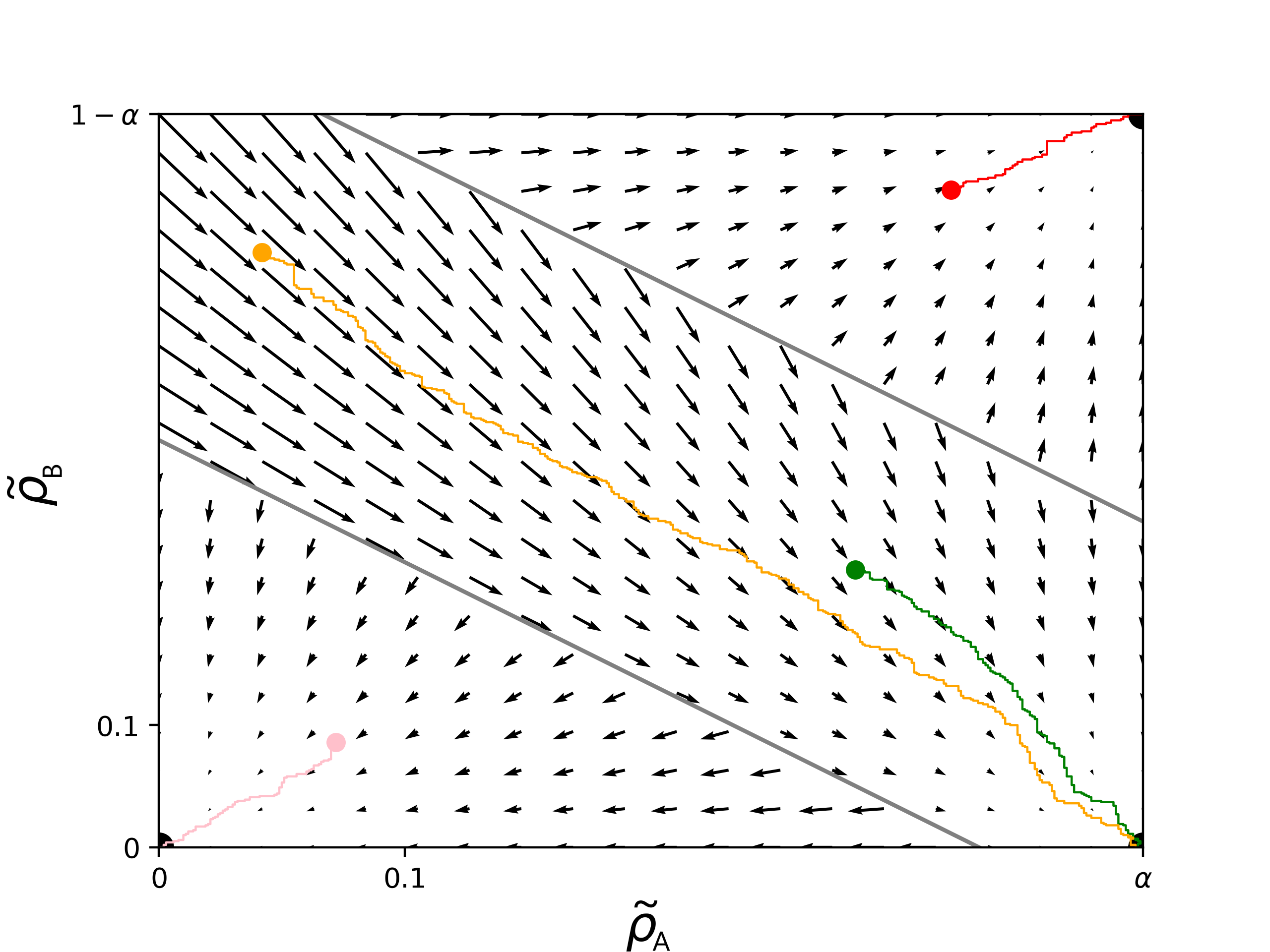}}
  \hfill  
  \subfigure[$\;k = 300,\; S=0.8$]{\includegraphics[width=0.42\textwidth]{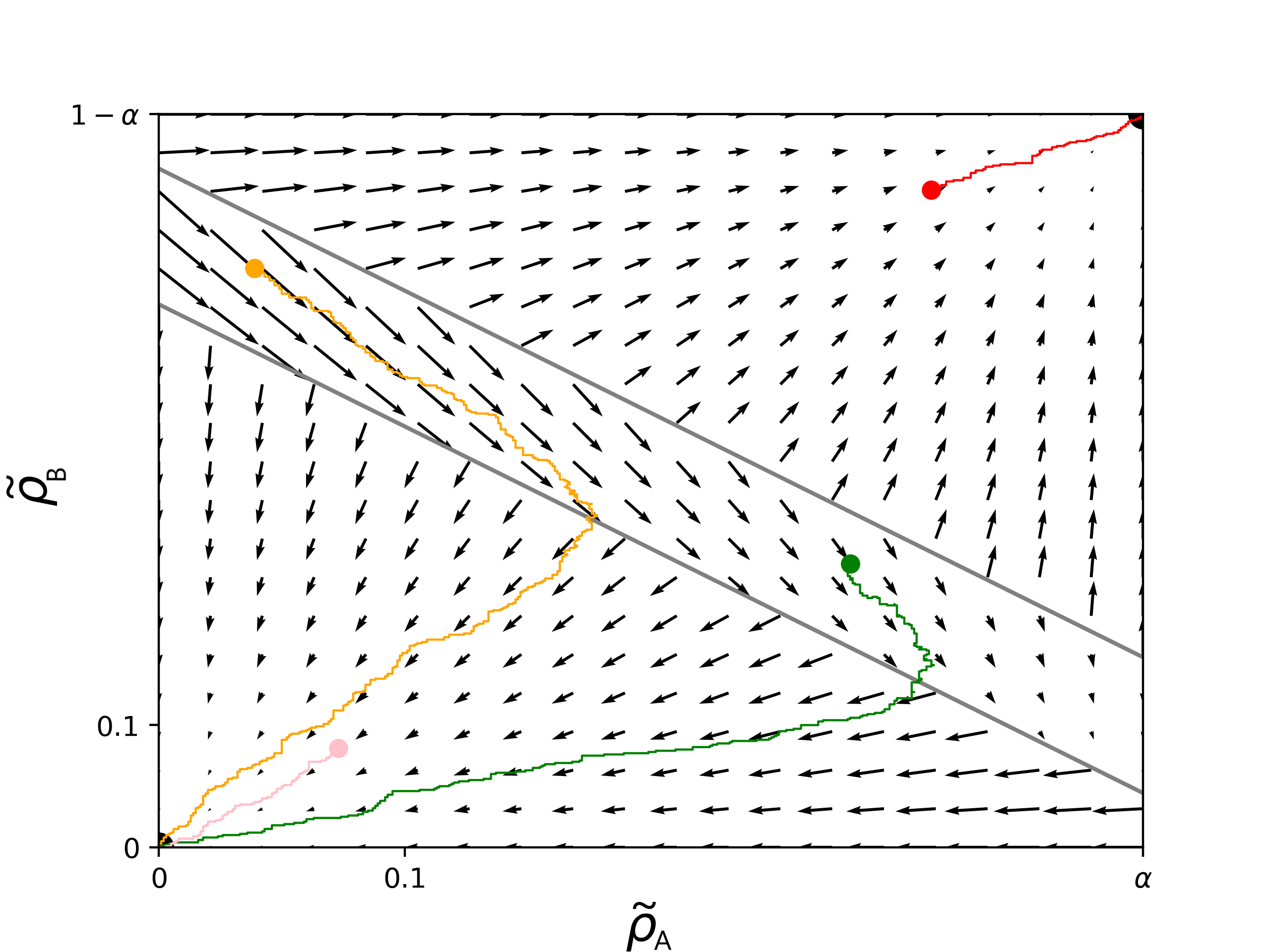}}
  \vfill
  \subfigure[$\;k = 30,\; S=0.8$]{\includegraphics[width=0.42\textwidth]{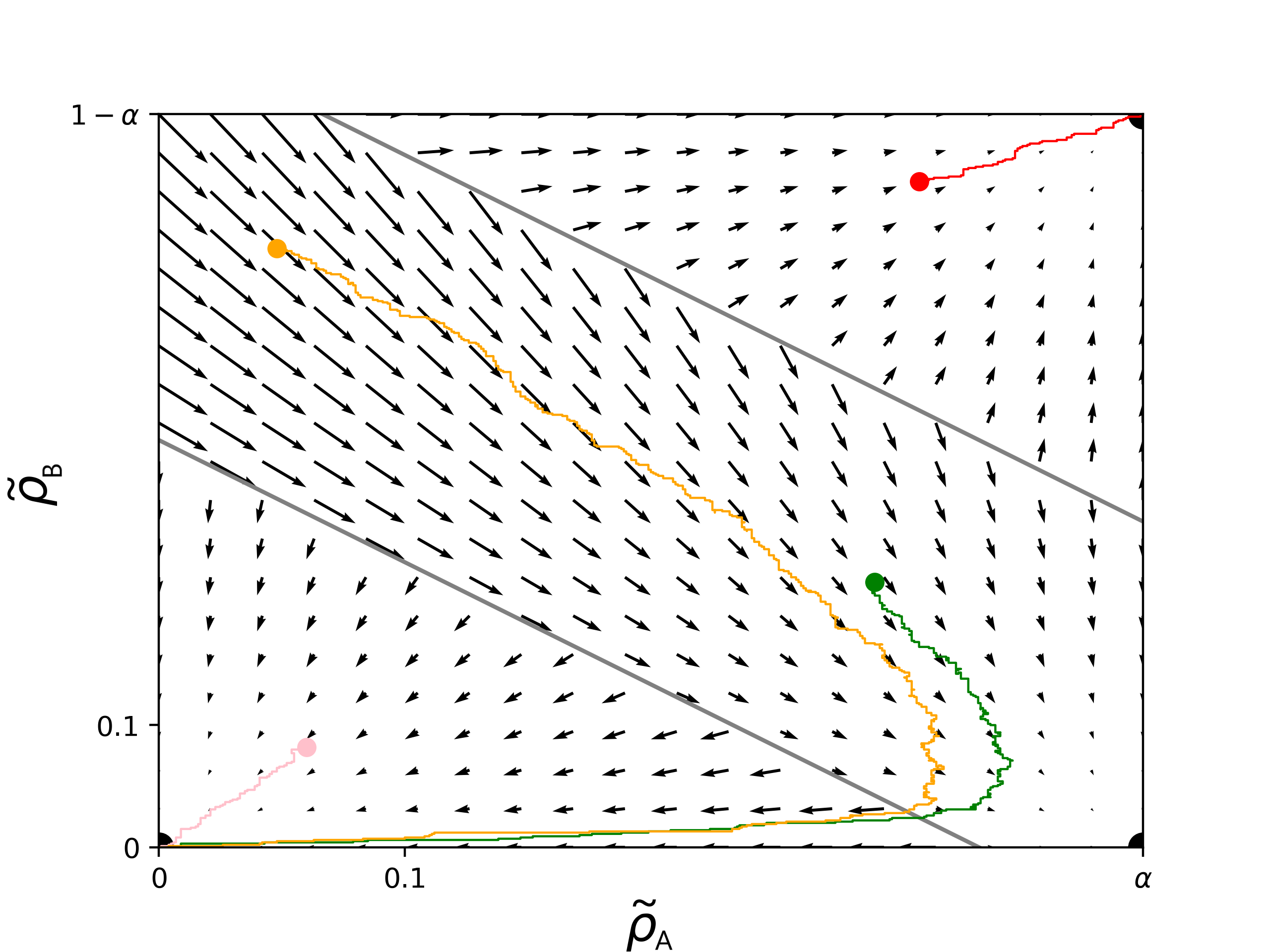}}
  \hfill
  \subfigure[]{\includegraphics[width=0.42\textwidth]{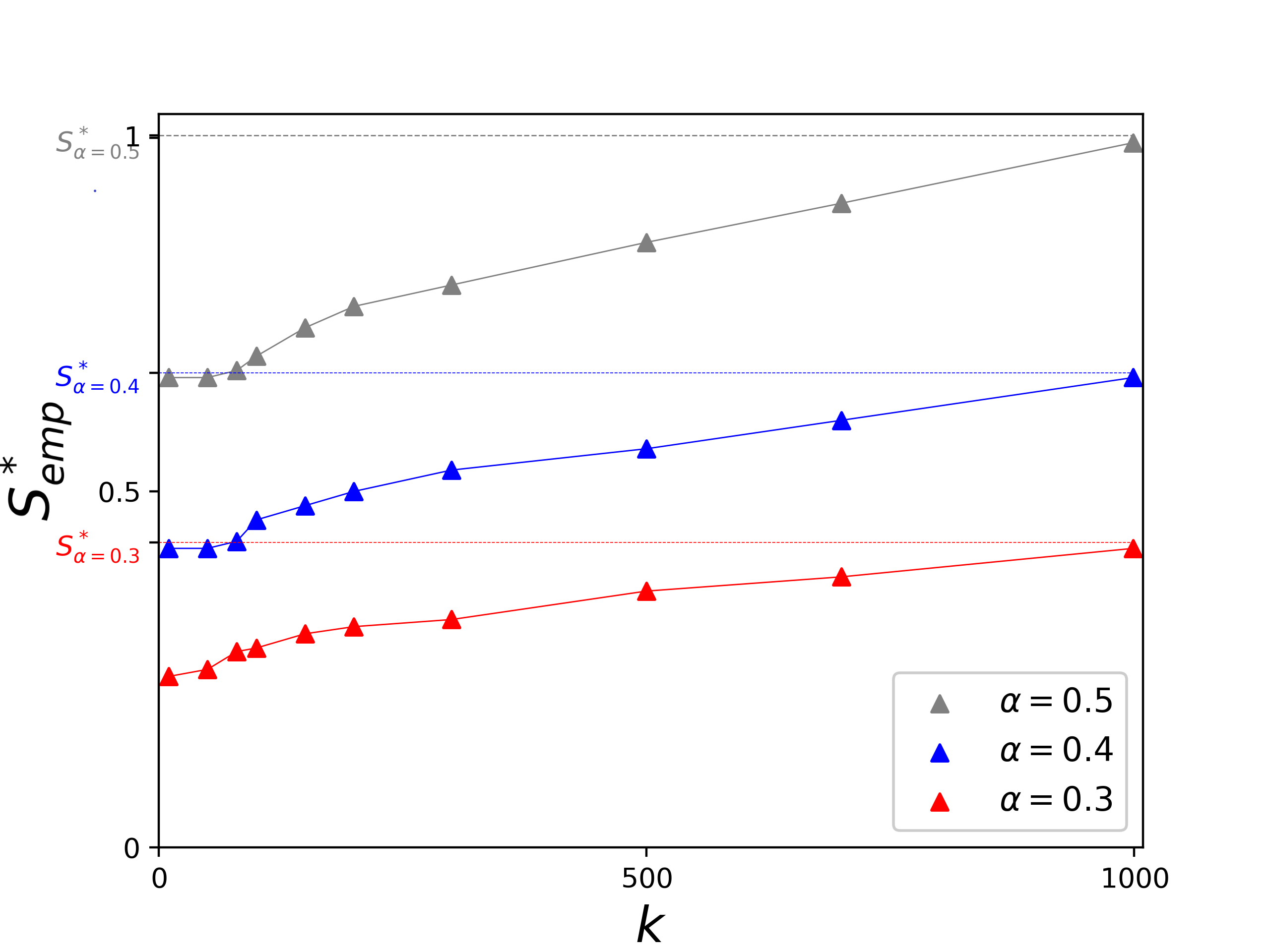}}
  \caption{\textbf{Best-response: mean-field predictions and low degree effects.}  \textit{(a),(b)} and \textit{(c)} show the vector field related to the mean-field predictions of the dynamics (eq. \ref{eq best response dynamics}) and the corresponding stable fixed points (black dots at the corners). Moreover, the trajectories of the system evolutions from various initial conditions, corresponding to different colors, are reported. The initial states are indicated with the colored dots.
  In all the simulations and predictions $N=1000$ and $\alpha = 0.4$. In \textit{(a)} and \textit{(c)}, as $S<S^*$ the polarized state is a stable fixed point. Nevertheless, the polarized state is actually reached only in \textit{(a)} ($k=300$), while for a lower degree, in \textit{(c)} ($k=30$), it is never reached as the mean-field assumptions break. 
  Figure \textit{(d)} reports the empirical thresholds (triangles) as a function of the degree of the regular graph, for various values of $\alpha$ as in the legend and $N=1000$. For each $\alpha$, the mean-field prediction $S^*_\alpha$ \eqref{mf threshold best response} is also reported (horizontal thin lines).   }
  \label{fig best response}
\end{figure}

\noindent 
Here, we analyze the case in which all the selected agents choose the action that gives them the highest payoff. This dynamical rule (best response) corresponds to assuming agents with infinite rationality ($R\rightarrow\infty$). 
We write the dynamical equations in the mean-field (well-mixed) approximation and prove that, in this case, only two or three equilibrium states are present: two consensus states at one or the other strategy, and one polarized state where each individual chooses its preferred opinion. We find that the polarized state is a (stable) fixed point of the system only for some combinations of the preference intensity $S$ and the system composition $\alpha$. \\
\\
We refer to $\alpha\equiv \alpha_A$ as the fraction of individuals with $1$ as preferred action (and thus $\alpha_B= 1-\alpha$), as in the main text, and to
\begin{equation}
    \tilde{\rho}_A = \frac{\sum_{i:\theta_i=1}{x_i}}{N}\in [0,\alpha] \quad \quad \quad  \tilde{\rho}_B = \frac{\sum_{i:\theta_i=0}{x_i}}{N} \in [0,1-\alpha]
\end{equation}
\noindent as the time-dependent number of agents with preferences respectively $1$ and $0$ playing action $1$, over the total number of agents. Notice that $\tilde{\rho}_{A/B} = \alpha_{A/B}\rho_{A/B} $.\\
It is easy to see that the preference intensity determines the threshold of nearest neighbours playing action $+1$ over which is convenient to play $+1$ (without any preference this threshold is $\frac{1}{2}$) and viceversa. Within the well-mixed approximation, i.e. when each agent's ego-network perfectly reflects the whole system and no local fluctuations are included, the changes $\dot{\tilde{\rho}}_A,\dot{\tilde{\rho}}_B$ of the dynamical variables per unit time (1 unit time corresponds to $N$ steps of the dynamics, in order to account for the system size) read
\begin{equation}
    \begin{cases}
        \dot{\tilde{\rho}}_A = (\alpha-\tilde{\rho}_A)\theta(\tilde{\rho}_A+\tilde{\rho}_B- \frac{S}{1+S}) - \tilde{\rho}_A\theta(\frac{S}{1+S} - (\tilde{\rho}_A+\tilde{\rho}_B))\\
        \dot{\tilde{\rho}}_B = (1-\alpha-\tilde{\rho}_B)\theta(\tilde{\rho}_A+\tilde{\rho}_B- \frac{1}{1+S}) - \tilde{\rho}_B\theta(\frac{1}{1+S} - (\tilde{\rho}_A+\tilde{\rho}_B))
    \end{cases}
\label{eq best response dynamics}
\end{equation}
where $\theta(x)$ is the Heaviside step function, i.e. equal to $1$ if $x>0$ and $0$ otherwise. Thus the dynamics can be divided into three zones, depending on the total density of agents playing $1$ in the system $\rho = \rho_A + \rho_B = \frac{\sum_{i}{x_i}}{N}$:\\
when $\rho < \frac{S}{1+S}$
\begin{equation}
    \begin{cases}
        \dot{\tilde{\rho}}_A =  - \tilde{\rho}_A \\
        \dot{\tilde{\rho}}_B =  - \tilde{\rho}_B 
    \end{cases}
\end{equation}
when $\rho \in [\frac{S}{1+S},\frac{1}{1+S}]$
\begin{equation}
    \begin{cases}
        \dot{\tilde{\rho}}_A =  \alpha - \tilde{\rho}_A  \\
        \dot{\tilde{\rho}}_B =  - \tilde{\rho}_B  
    \end{cases}
\end{equation}
and when $\rho > \frac{1}{1+S}$
\begin{equation}
    \begin{cases}
        \dot{\tilde{\rho}}_A =  \alpha - \tilde{\rho}_A  \\
        \dot{\tilde{\rho}}_B =  1- \alpha - \tilde{\rho}_B 
    \end{cases}
\end{equation}
It is easy to verify, by solving the fixed point equations and performing a simple linear stability analysis, that each of the zones potentially has a stable fixed point: these are, in the $\rho_A,\rho_B$ plane, respectively $(0,0)$ for the first,$(\alpha,0)$ for the second and $(\alpha,1-\alpha)$ for the third zone. If it is true that both the consensus state are always located within the corresponding areas and thus always exist, this cannot be said for the fully polarized one: $(\alpha,0)$ is a (stable) fixed point if and only if it falls in the area defined by the second condition $\rho \in [\frac{S}{1+S},\frac{1}{1+S}]$, so if $\alpha$  is in the range $[\frac{S}{1+S},\frac{1}{1+S}]$. If without loss of generality, we take $\alpha<0.5$, the latter condition corresponds to a threshold on the preference intensity 
\begin{equation}
     S^* = \frac{\alpha}{1-\alpha}
\label{mf threshold best response}
\end{equation}
For values of $S$ above $S^*$, starting from everyone choosing his preferred action the system will move to consensus towards one of the two opinions (typically the one of the majority), while for $S<S^*$ the preference intensity is sufficient to make the system remain in the fully polarized state. The mean-field dynamics is represented by the vector field and it is tested on a large degree network $k=300$ of $N=1000$ agents for multiple initial conditions, in figure \ref{fig best response}a for $S=0.5$ and in figure \ref{fig best response}b for $S=0.8$. \\
When testing the mean-field predictions on a sparser graph, we see that the accuracy of the mean-field predictions and specifically the mean-field threshold (\ref{mf threshold best response}) considerably decreases even for random graphs with large average degrees, as shown in figure \ref{fig best response}c for $k=30$. In figure \ref{fig best response}d we report the empirical threshold as a function of the average degree of a random regular graph of $N=1000$ agents, for various compositions $\alpha$ corresponding to the different colours. The empirical threshold is defined as the largest value of $S$ for which, starting from the fully polarized state of the system, over an ensemble of the system's trajectories the majority of them do not approach full consensus. The motivation for the correction to the mean-field predictions for the threshold resides in the validity of the mean-field assumption that assumes that each agent's ego-network is a perfect sample of the whole network, both topologically (the neighbours' belonging classes) and dynamically (distribution of current opinions): in a random graph some nodes have an ego-network that deviates from the average one, the more the smaller is the average degree. Thus, having e.g. a higher number of connections with one class with respect to the average, the agents would adopt more easily that class' preference and may induce other agents of the same population to follow them, generating a cascade effect that provokes consistent fluctuations, i.e. deviations from the mean-field predicted behaviour, possibly making the system fall into the basin of attraction of the consensus points. By analysing the numerical simulations, we speculate that the effect is amplified by the noise induced by bounded rationality, which generally increases the fluctuations and thus the possibility of falling into the basin of attraction of the equilibrium states corresponding to (almost-)consensus.

\section*{Acknowledgement}
MAJ is supported by the PNRR NQST (Code: $PE23$).

\end{document}